\documentclass[12pt,preprint]{aastex}

\newcommand{\be}{\begin{equation}}
\newcommand{\ee}{\end{equation}}
\newcommand{\ba}{\begin{eqnarray}}
\newcommand{\ea}{\end{eqnarray}}

\slugcomment{Accepted to Astrophysical Journal}
\begin{document}

\title{Velocity Modification of Power Spectrum from
Absorbing Medium}

\author{A. Lazarian}
\affil{Department of Astronomy, University of Wisconsin,
Madison, US}
\author{D. Pogosyan}
\affil{Physics Department, University of
Alberta, Edmonton, Canada}

\begin{abstract}
Quantitative description of the statistics of intensity fluctuations
within spectral line data cubes introduced in our earlier work
is extended to the absorbing media. A possibility of extracting
3D velocity and density statistics from both integrated line
intensity as well as from the individual channel maps is analyzed.
 We find that absorption enables the velocity effects to be seen
even if the spectral line is integrated over frequencies. This
regime that is frequently employed in observations
is characterized by a non-trivial
relation between the spectral index of velocities and the spectral
index of intensity fluctuations. 
For instance when density is dominated by fluctuations
at large scales, i.e. when correlations scale as $r^{-\gamma}, ~\gamma<0$,
the intensity fluctuations exhibit a
universal spectrum of fluctuations $\sim K^{-3}$ over a range of scales. When
small scale fluctuations of density contain most of the energy,
i.e. when correlations scale as $r^{-\gamma}, ~\gamma>0$, 
the resulting spectrum  of the integrated lines
depends on the scaling of the underlying density
and scales as $K^{-3+\gamma}$. We show that if we take the spectral
line slices that are sufficiently thin we recover our earlier results for
thin slice data without absorption. 
As the result we extend the  Velocity Channel Analysis (VCA)  
technique  to optically thick lines enabling studies of turbulence in molecular
clouds. In addition, the developed mathematical machinery enables a
quantitative approach to solving other problems that involved statistical
description of turbulence within emitting and absorbing gas.
      
\end{abstract}


\section{Introduction}

There is little doubt that interstellar medium is turbulent
(see reviews by Armstrong, Spangler \& Rickett 1995; 
Lazarian 1999a, Lazarian, Pogosyan \& Esquivel 2002).
Turbulence proved to be ubiquitous in molecular clouds (Dickman 1985),
diffuse ionized (Cordes 1999) and neutral (Lazarian 1999b)
media. Magneto-hydrodynamical (MHD) turbulence controls
many essential astrophysical processes including star formation, transport
and acceleration of cosmic rays, of heat and mass (see Schlickeiser 1999,
Vazquez-Semadeni et al 2000, Narayan \& Medvedev 2001, Wolfire et al. 2002, Cho et al.
2003, see recent reviews
by Cho, Lazarian \& Vishniac 2002, henceforth CLV02a, Elmegreen \& Scalo 2004, Lazarian \& Cho 2004).  
Unfortunately, we are still groping for the
basic properties of interstellar turbulence. For instance,
it is unclear whether turbulence in molecular clouds is a part of 
a global turbulent cascade or has its local origin. 

Recent years were marked by a
substantial progress in theoretical and numerical description of both
incompressible and compressible MHD turbulent
cascade (Goldreich \& Shridhar 1995, 
Lithwick \& Goldreich 2001, Cho \& Lazarian
2002a, CLV02a, Cho \& Lazarian 2003ab). It is encouraging 
that the measured statistics
of electron density fluctuations (see Armstrong et al. 1995),
synchrotron and polarization degree fluctuations (see Cho \& Lazarian
2002b) are in rough correspondence with the theoretical expectations
(see more discussion in Cho \& Lazarian 2003b). Further testing
requires better input data, and this makes the
determination of actual power spectra of interstellar
turbulence extremely important and
meaningful. Indeed, it gives a chance to distinguish between different
pictures of turbulence, e.g. shocks versus eddies picture,
 and put theoretical constructions developed for various
interstellar processes, e.g. cosmic ray propagation (see
Yan \& Lazarian 2002), grain dynamics (Lazarian \& Yan 2002, Yan \&
Lazarian 2003), stochastic reconnection (Lazarian \&
Vishniac 1999, Lazarian, Vishniac \& Cho 2003) on a solid ground.
Even within the picture of MHD cascade that is advocated by the simulations
in Cho \& Lazarian (2002) and which does not contain shocks, the distribution
of energy between Alfven, slow and fast modes does depend on the turbulence
driving. Therefore the spectra measured from observations can be used 
to constrain the driving.
In addition, the particular features of the power spectrum,
which are associated with energy injection and dissipation are important.
Identification of such features and the spatial variations of their
properties will shade light to the dynamics of interstellar gas and processes
of star formation (see Lazarian \& Cho 2003). 

We should stress that while the statistics of density fluctuations can
be only considered as an indirect way of testing turbulence, the
 spectral surveys contain the information
about the direct measure, namely, the velocity statistics.
This statistics is extremely valuable, provided that we can extract
it from the observational data. 
As MHD turbulence is indeed an interdisciplinary subject, high resolution
studies of interstellar turbulence can test contemporary ideas of
MHD cascade. The implications of the improved understanding would
affect description of wide range of astrophysical
phenomena from solar flares to gamma ray
bursts (see Lazarian et al. 2003). 

Additional motivation for studying interstellar turbulence stems from
the advances in direct numerical
simulations of interstellar medium (see review by Vazquez-Semadeni
et al 2000). Unlike  MHD simulations focused on obtaining
fundamental properties of turbulence (see CLV02a) 
those of interstellar medium attempt to include various aspects of
interstellar physics. This usually limits the inertial
range covered. However, testing of the
numerical results against observations is
becoming more interesting as numerical resolution increases. With other
techniques of comparing numerics with observations having their
problems (see review by Ostriker 2002, also Brunt et al. 2003) power spectra 
and second order
correlation functions look as a promising tool for the future (see
Lazarian 1995, Lazarian \& Pogosyan 2000, henceforth LP00,
Vestuto, Ostriker \& Stone 2003). The potential of combining different
techniques in order to get the properties of turbulence is also
large.

Although in this paper we primarily refer to interstellar processes,
the technique we discuss has a broader application. For instance,
recent advances in X-ray astronomy have allowed to get information on
turbulence in intracluster gas (see Inogamov \& Sunyaev 2003,
Sunyaev, Norman \& Bryan 2003). The issues that researchers face
there are similar to those dealt in interstellar turbulence studies.

To get insight into turbulent processes statistical approach  is
useful (see Monin \& Yaglom 1976, Dickey 1995). So far, in astrophysical
context the most tangible progress was achieved
via scintillations and scattering technique (see Spangler 1999). Those 
measurements are limited to probing fluctuations  
of electron density on rather small scales,
namely, $10^8$--$10^{15}$~cm. This research profited a lot from an adequate
theoretical understanding of processes of scintillations and scattering
(Goodman \& Narayan 1985) and this made it very different from other
branches of interstellar turbulence research.  
  In comparison, formation of emission line profiles in turbulent media
has not been properly described until very recently and it is natural
that numerous attempts to study turbulence in diffuse interstellar
medium, HII regions and molecular clouds using emission lines
(see Munch 1958, O'Dell 1986, O'Dell \& Castaneda 1987,
Miesh \& Bally 1994, reviews by Scalo 1987, Lazarian 1992) 
were only partially successful. This is very
unfortunate as the line profiles contain unique information about
turbulent velocity field. We reiterate that the studies of stochastic
density provide only indirect insight into turbulence and cannot
distinguish between active and fossil turbulence pictures. 
 
Studies of the velocity field have been attempted at different
times with velocity centroids
(e.g. Munch 1958, Miesh, Bally \& Scalo 1999,  Miville-Deschenes, Levrier \& 
Falgarone 2003). 
However, it has long been realized that the centroids are
affected, in general, by both velocity and density fluctuations
(see Stenholm 1989). An important criterion when centroids indeed
reflect the velocity statistics was obtained in Lazarian
\& Esquivel (2003, henceforth LE03). LE03 showed
that when the criterion is not satisfied the centroids
are dominated by density. LE03 pointed out that the velocity centroids  
can be modified to extract the velocity contribution (LE03), but the 
resulting measures may be pretty noisy. As the result, velocity
centroids taken alone may not be sufficiently robust and reliable technique.
A recent study in Esquivel \& Lazarian (2004) revealed that for Mach numbers
larger than 2 or 3 the traditional velocity centroids fail. As Mach numbers
as high as 10 are expected for molecular clouds the applicability of
the traditional centroids to molecular clouds is highly questionable. 

Principal Component Analysis (PCA) 
of the emission data (Heyer \& Schloerb 1997,
Brunt \& Heyer 2002ab) has been suggested as a new way to study 
interstellar turbulence. However, recent testing showed
that it provides statistics different from power spectra (Heyer, private
communication, Brunt et al 2004). What sort of statistics we can 
obtain using the PCA requires a further study.

Wavelet analysis (see Gill \& Henriksen 1990), 
Spectral Correlation Functions\footnote{Spectral Correlation Functions
if they are measured for the same velocity present just another way
of dealing with fluctuations within channel maps. Therefore all the
VCA results are applicable to them. If one tries to generalize 
Spectral Correlation Functions to the velocity direction (see Lazarian 1999)
only limited information about turbulence can be available using
the tool (see Appendix C).} (see Rosolowsky et al. 1999,
Padoan, Rosolowsky \& Goodman 2001),
genus analysis (see Lazarian, Pogosyan \& Esquivel 2002) are other
important statistical tools. They can provide statistics and insight 
complementary to power spectra (see review by Lazarian 1999a).   
Synergy of different techniques should provide the necessary
insight into turbulence and enable the comparison of observational statistics
with theoretical expectations (see Cho \& Lazarian 2003a for a
review).

Velocity statistics contributes to intensity variations observed in
channel maps. Power-law spectra suggestive of underlying turbulence were
obtained at different times by a number of researchers (see Kalberla
\& Mebold 1983,
Green 1993). 
The problem that plagued those
studies was inability to separate contributions due to velocity
and density. Indeed, both velocity and density fluctuations 
affect the small-scale emissivity fluctuations
observed at a given velocity. Therefore the separation of 
 contributions to velocity and density requires a quantitative
description of spectral line data cube statistics.

This problem was addressed in Lazarian
\& Pogosyan (2000, henthforth LP00), where the fluctuations
of intensity in channel maps were related to the statistics of velocity 
and density.  
LP00 introduced a new statistical technique that was termed in
Lazarian, Pogosyan \& Esquivel (2002)
Velocity Channel Analysis (henceforth VCA). Within VCA 
the separation of velocity and density contributions is obtained
by changing the thickness of the analyzed slice of the Position-Position-
Velocity (PPV) data cube. The theoretical description of the intensity
statistics for both thick and thin velocity slices enabled LP00 to
disentangle velocity and density contributions to  
 intensity fluctuations in PPV cubes.  

The VCA was successfully tested numerically in Lazarian et al. (2001)
and Esquivel et al. (2003). Those tests employed the velocity
and density obtained via simulations of compressible MHD turbulence
to obtain synthetic maps, which were analyzed using VCA to recover
underlying velocity statistics.
Since then the VCA has been successfully applied to get velocity
spectra of turbulence (see section~5).

The VCA in its existing form, however, does not account for absorption
of radiation. LP00 argued that for the 21~cm emission 
data the effect of absorption is marginal in the direction of the galactic
anti-center for which the technique was applied.
However, the data in Dickey et al
(2001) show substantial absorption in HI towards inner part of the Galaxy.
Moreover, the absorption is definitely essential for other, e.g. CO 
transitions.

The purpose of this paper is to extend the earlier obtained theoretical description of the
Doppler-shifted spectral line data cubes
by including the treatment of absorption effects and thus to obtain a
quantitative tool to study turbulence in various interstellar conditions, 
including molecular clouds. We introduce the 3D anisotropic
statistics of fluctuations within PPV data cubes in section 2, describe
effects absorption in section~3, provide the statistics for thin
and thick slices in section~4, explain some of our findings in physical
terms in section~5, discuss our results and their implications
for observational data in section~6. The summary is provided in
section~7. The 
lengthy derivations are collected in the appendixes which
are an important part of the paper. 

\section{Statistics of Velocity and Density}

\subsection{Basics of Turbulence Statistics}

{\it Statistics in Real Space}\\
In the presence of magnetic field MHD turbulence gets axisymmetric
in the system of reference related to the {\it local} direction
of magnetic field. In this system of reference magnetic fields can
be easily mixed by turbulence in the direction perpendicular to magnetic
fields. Those mixing motions generate wave-like perturbations 
 propagating along magnetic field lines. As the kinetic energy decreases
with the decrease of the scale, the weak mixing motions bend magnetic
field lines less and less and the eddies get more and more elongated.

The Goldreich \& Shridhar (1995) model of incompressible turbulence 
prescribes Kolmogorov scaling of mixing motions perpendicular to magnetic
field lines (see a simplified discussion in CLV02a) and the elongation of
the velocity fluctuations that increases as $l^{1/3}$.
Further research in Lithwick \& Goldreich (2001) 
and Cho \& Lazarian (2002) has shown that the 
basic features of the Goldreich-Shridhar turbulence carry on 
for Alfvenic perturbations to the
compressible regime\footnote{Magnetosonic fast modes that  arise in
compressible fluid, are, however, isotropic (Cho \& Lazarian 2002, 2003ab).
 This entails many important astrophysical
consequences (see review
by Lazarian, Cho \& Yan 2002). When both fast and Alfven modes
are present in the fluid, the total anisotropy of magnetic and velocity
fluctuations decreases.}.

Observations, however, are usually unable to identify the
local orientation of magnetic field and deal with
the magnetic field projection integrated over the line of sight.
As the result the locally defined perpendicular and parallel directions
are mixed together in the process of observations (see
CLV02a). The fluctuations
with more power dominate the signal for both the direction of
the averaged ${\bf B}$ and perpendicular to it. There is some
residual anisotropy, but this anisotropy is scale-independent 
and is determined by the rate of the meandering of the large scale 
field. 
In other words, from the
observational point of view the spectra of intensity fluctuations 
obtained 
from the Goldreich \& Shridhar turbulence and the isotropic 
turbulence are similar. Observations in Green (1993)
and numerical studies in Esquivel et al. (2003)
support this picture.   

The facts above allow for a substantial simplification of the turbulence
description, namely, they permit us to use standard isotropic statistics
(Monin \& Yaglom 1976).
Obtaining this statistics would correspond to obtaining the scaling
of density and velocity fluctuations perpendicular to magnetic field lines. 

Statistically isotropic in $xyz$-space density field $\rho({\bf x})$
has the correlation function
\be
\xi(r)=\xi({\bf r}) = \langle \rho ({\bf x}) \rho ({\bf x}+{\bf r}) \rangle~~
~.
\label{xifirst}
\ee
We shall also use the correlation function of density fluctuations
$\delta\rho=\rho-\langle\rho\rangle$
\be
\tilde\xi(r)= \langle \delta\rho ({\bf x})
\delta\rho ({\bf x}+{\bf r})\rangle = \xi(r)-\langle\rho\rangle^2
~.
\label{tildexi}
\ee

The structure function 
\be
d(r)=
\langle(\rho({\bf x}+{\bf r})-\rho({\bf x}))^2 \rangle~~~,
\label{2}
\ee
is another way of describing turbulence. Here $\tilde d(r)$ for fluctuations
coincides with $d(r)$.

Statistical descriptors of fluctuations can often be assumed to have 
power-law dependence on scale $\propto r^{-\gamma}$.
If the correlation is dominated by large scales, the structure functions
are used and $\gamma < 0$, while for small scale dominated statistics
the correlation functions are usually used and $\gamma > 0$.
We show in Appendix~A that the two cases in the
asymptotic regime of small $r$ can be treated very similarly. Therefore
we will frequently refer to correlation functions having in mind both
$\gamma>0$ and $\gamma<0$.
 
An isotropic velocity field ${\bf u}({\bf x})$ is fully described by
the structure tensor $\langle \Delta u_i \Delta u_j \rangle$,
which can be expressed via longitudinal
$D_{LL}$ and transverse $D_{NN}$ components
(Monin \& Yaglom 1972)
\begin{equation}
\langle \Delta u_i \Delta u_j \rangle = \left( D_{LL}(r)-D_{NN}(r) \right) {r_i r_j \over r^2}
+D_{NN}(r) \delta_{ik}~~~,
\label{struc}
\end{equation}
where $\delta_{ik}$ equals 1 for $i=k$ and zero otherwise. 
We define z-projection of the velocity structure function as
\begin{equation}
D_z({\bf r}) \equiv \langle \Delta u_i \Delta u_j \rangle \hat z_i 
\hat z_j = D_{NN}(r) + [ D_{LL}(r)-D_{NN}(r)] \cos^2\theta~,
~~~ \cos\theta \equiv {\bf \hat r \cdot \hat z}
\label{eq:Dz}
\end{equation}
which for the power-law velocity gives
\be
D_z({\bf r})=Cr^{m}[1+\frac{m}{2}(1-\cos^2\theta)]~~~,
\ee
if we assume that the velocity is solenoidal\footnote{Separating
solenoidal and potential components of the velocity is an important
problem that we do not address here. It was suggested in LE03 that this separation
could be done by combining
VCA and velocity centroids.}.
For this paper it is only important that
$D_z\sim Cr^m$. A discussion of the shallow and steep spectra 
of density as well as of the
velocity statistics within finite size clouds is given in Appendix A.

{\it Spectra and Correlation Functions}\\
Spectra and correlation/structure functions are two complementary ways
of describing turbulence. Given an N-dimensional correlation function
$\xi_N(\bf r)$
one can obtain the spectrum
\be 
P({\bf k})=\int d^N {\bf r} e^{i{\bf k}{\bf r}} \xi_N(\bf r)~~~,
\label{spectrum:gen}
\ee
where the integration is performed in the $N$-dimensional space.

In LP00 the Fourier space statistics, namely, 
spectra were widely used with the correlation functions playing
an auxiliary role. In the present paper we 
deal with absorption defined in real space. Therefore we use
real space statistics, namely,  correlation and structure functions  
functions. It is obvious from eq.~(\ref{spectrum:gen}) that for power-law 
correlation function (CF), i.e. $\xi_N\sim r^{-\gamma}$ the spectral index is
also a power-law, i.e. $P\sim k^{n}$, where $n=-N+\gamma$. In other words, 
\be
({spectral~index})= (-{dimensions~of~space}-{CF~index})
\ee    
In Kolmogorov turbulence passive scalar density correlations scale as
$r^{2/3}$, which corresponds to $\gamma=-2/3$ in our notations. Thus
the Kolmogorov spectrum index is $-11/3$. In turbulence literature
$E(k)=4\pi k^2 P(k)$ is usually used.  In this notation the Kolmogorov spectrum
is  $E(k)\sim k^{-5/3}$, and thus, the Kolmogorov law is often referred to as $-5/3$
law.

\subsection{Statistics in PPV}
One does not observe the gas distribution in the real space galactic
coordinates xyz where the 3D vector ${\bf x}$ is defined.
Rather, intensity of the emission in a given spectral
line is defined in Position-Position-Velocity (PPV) cubes towards some direction on the sky and at a
given line-of-sight velocity $v$.
\footnote{All velocities are
the line-of-sight velocities, we omit any special notation to denote
z-component.} 
In the plane parallel approximation the direction on the sky is identified
with xy plane where the 2D  spatial vector ${\bf X}$ is defined,
so that the coordinates of PPV cubes available through observations
are $({\bf X},v)$.
The relation between the 
real space and PPV descriptions is defined by a map $({\bf X},z)
\to ({\bf X},v)$. 

The central object for our study is a turbulent cloud
in PPV coordinates. The statistical properties of
the PPV density  $\rho_s({\bf X},v)$ 
depend on the density of gas in real galactic coordinates,
but also on velocity distribution of gas particles.
Henthforth we use the subscript $s$ to distinguish 
the quantities in $({\bf X}, v)$ coordinates from those $({\bf X}, z)$
 coordinates.
We shall always assume 2D statistical homogeneity and isotropy
of $\rho_s({\bf X},v)$
in ${\bf X}$-direction over the image of a cloud.
 However, homogeneity along the velocity direction can only be assumed after
additional considerations, if at all.
Naturally, there is no symmetry between $v$ and ${\bf X}$.

We shall use the following notation for statistical descriptors of the
density in PPV space: the mean density is
$\bar\rho_s(v_1)=\langle \rho_s({\bf X}_1,v_1) \rangle$,  
and the correlation functions of the density and, closely related, the density
fluctuations $
\delta\rho_s({\bf X}_1,v_1) = \rho_s({\bf X}_1,v_1) - \bar \rho_s(v_1) $
are
\begin{eqnarray}
\xi_s(R,v_1,v_2) &\equiv& \langle \rho_s({\bf X}_1,v_1)
\rho_s({\bf X}_2,v_2) \rangle \\
\tilde \xi_s(R,v_1,v_2) &\equiv& \langle \delta\rho_s({\bf X}_1,v_1)
\delta\rho_s({\bf X}_2,v_2) \rangle = 
\xi(R,v_1,v_2) - \bar \rho_s(v_1) \bar\rho_s(v_2) ~.
\end{eqnarray}
Here we have maintained notation which highlights the
symmetries of the statistics. Homogeneity and isotropy in 
Position-Position coordinates
${\bf X}$ lead the mean density to depend only on velocity while
 the correlation
function depends only on the magnitude of separation between two sky directions
$R=|{\bf R}|=|{\bf X}_1-{\bf X_2}|$. Dependence on the velocity is retained
for the time being in the general form.

For PPV statistics it is more convenient to use 
the structure functions
\begin{eqnarray}
d_s(R,v_1,v_2)&=&\langle \left[ \rho_s({\bf X}_1,v_1)-\rho_s({\bf X}_2,v_2)\right]^2\rangle \\
\tilde d_s(R,v_1,v_2)&=&\langle \left[\delta\rho_s({\bf X}_1,v_1) -
\delta\rho_s({\bf X}_2,v_2)\right]^2 \rangle 
= d_s(R,v_1,v_2) - \left[ \bar\rho_s(v_1) - \bar\rho_s(v_2) \right]^2
\end{eqnarray}

Let us derive the two-point correlation function in PPV space without
using the power spectrum formalism employed in LP00.  
The line-of-sight component of velocity $v$ 
at the position ${\bf x}$ is a sum of the 
regular gas flow (e.g., due to galactic rotation) $v_{gal}({\bf x})$,
the turbulent velocity $u({\bf x})$ and the residual component
due to thermal motions. This residual thermal
velocity $v-v_{gal}({\bf x})-u({\bf x})$ has a Maxwellian distribution
\be
\phi_v({\bf x}) {\mathrm d} v =\frac{1}{(2\pi \beta)^{1/2}}
\exp\left[-\frac{(v-v_{gal}({\bf x})-u({\bf x}))^2}
{2 \beta }\right] {\mathrm d} v ~~~,
\label{phi}
\ee
where $\beta=\kappa_B T /m$, $m$ being the mass of atoms.
The temperature $T$ can, in general, vary from point to point. 

The density of the gas in the PPV space 
$\rho_s({\bf X},v)$ can then be written as
\begin{equation}
\rho_s({\bf X},v){\mathrm d {\bf X} d} v = 
\left[ \int_0^S {\mathrm d} z\; \rho({\bf x}) \phi_v({\bf x}) \right]
{\mathrm d {\bf X} d} v  ~~~,
\label{rhoz}
\end{equation} 
where $\rho({\bf x})$ is the (random) density of gas in real space.
This expression just counts the number of atoms
along the line-of-sight that have z-component of velocity in the interval
$[v,v+dv]$.  The limits of integration are defined by the spatial extend $S$ 
of the emitting gas distribution.

To compute the 
correlation function $\xi_s$
we disregard the effect of correlations between density and velocity
fluctuations\footnote{A similar
assumption was used in LP00. It was tested by Lazarian et al (2001)
and Esquivel et al. (2003), that
this assumption does not cause any problem velocity and density
are correlated.}:
\begin{eqnarray}
\langle \rho_s({\bf X}_1, v_1)
\rho_s({\bf X}_2,v_2)\rangle & = &
\int_0^S dz_1 \int_{0}^{S} dz_2 \;
\langle \rho({\bf x}_1)\rho({\bf x}_2)\rangle 
\;
\langle \phi_{v1}({\bf x}_1) \phi_{v2}({\bf x}_2) \rangle ~~~, \\
\langle \rho_s({\bf X},v) \rangle & = & \int_0^S dz
\langle \rho({\bf x}) \rangle \langle \phi_{v}({\bf x}) \rangle
\label{ksi1}
\end{eqnarray}
The brackets $\langle \cdots \rangle$ in eq.~(\ref{ksi1}) denote
statistical averaging over realizations of the random density $\rho({\bf x})$
and the turbulent velocity $u({\bf x})$ of gas. In particular, for the density
we can write $\langle \rho({\bf x}) \rangle = \bar \rho$ and 
$\langle \rho({\bf x}_1) \rho({\bf x}_2) \rangle = \xi (\bf r)$,
without any assumptions on density statistics.
For the Gaussian velocity field $u$, described by structure functions 
given by eq.~(\ref{struc}), the statistical average of the Maxwell functionals
(\ref{phi}) is given in Appendix~B (see (\ref{AppB:mean_cloud})
(\ref{AppB:2point_cloud}), (\ref{AppB:sigma})).

The general expression for the correlation function given
in the Appendix~B (see eq.(\ref{AppB:2point}) can be simplified if 
one neglects the edge effects associated with the wings\footnote{Edge effects
make the image of turbulence inhomogeneous. It is intuitively
clear that those effects should not affect the statistics
on the scale much less than the spectral line width.} of spectral lines (see
derivation of eq.~(\ref{AppB:2point_inf})).
In this case we have
\begin{equation}
\xi_s({\bf R},v) \sim
\frac{1} {2 \pi (C S)^{m/2}}
\int_{-S}^S {\mathrm d}z 
\int_{|z|/2}^{S-|z|/2} {\mathrm d}z_+ 
\; \xi({\bf r}) \; [D_z({\bf r})+2\beta]^{-1/2}
\exp\left[-\frac{(v-v_{gal})^2}{2 (D_z({\bf r})+2\beta)}\right],
\label{ksi2}
\end{equation}
where eq.~(\ref{struc_intro}) was used to write the prefactor explicitly. 
For a homogeneous statistics all correlation functions depend only on  
the separation between points.
Here and further on we
use variables without indexes to denote separation, as in ${\bf R=X_1-X_2}$,
$z=z_1-z_2$, $v=v_1-v_2$ and $v_{gal}=v_{gal,1}-v_{gal,2}$.
Variables with an index '+' denote the average of two arguments  
$z_+=(z_1+z_2)/2$, etc.
Dependence on these last quantities
appears only if statistical homogeneity is broken, e.g. by a finite
spatial extend of the gas distribution. 

In the absence  of the turbulent and thermal motions, $D_z,\beta \to 0 $,
the kernel in eq.~(\ref{ksi2}) reduces to the delta-function
$\delta(v-v_{gal})$,
and for monotonic $v_{gal}(z)$  
one can fully recover the line of sight 
position of an emitting atom
from its velocity. In a general case, 
while thermal broadening just smothers fluctuations out,
the effect of the  turbulent velocity fluctuations is scale dependent and
leads to a change of the correlation function slope. 

Equation~(\ref{ksi2}) allows for an arbitrary form of the regular
flow $v_{gal}$. A variable regular flow produces a 
complex map from galactic to velocity space by itself.
Two tractable cases are when the regular flow is of a simple linear form,
or the one when it is possible to attribute all motions to turbulence, thus setting $v_{gal}=0$.
In LP00 we have considered the case of linearized shear flow
$v_{gal}=f^{-1} z$ due to Galactic rotation:
\begin{equation}
\xi_s({\bf R},v) \sim
\int_{-\infty}^\infty {\mathrm d}z 
\; \xi({\bf r}) \; [D_z({\bf r})+2\beta]^{-1/2}
\exp\left[-\frac{(v-f^{-1}z)^2}{2 (D_z({\bf r})+2\beta)}\right],
\label{ksigal}
\end{equation}
where we have omitted an unimportant dimensional prefactor 
(see eq.~(\ref{AppB:2point_cloud}) for
the explicit form of the prefactor).
Galactic shear introduces a natural scale
$\lambda = {\left[f^2 C \right]}^{1 \over 2-m}$, 
at which the velocity dispersion $\sim C \lambda^{m}$
becomes equal to
the squared difference of the regular velocities determined by Galactic
rotation (i.e. $f^{-2}\lambda^2$). 
Asymptotics obtained in LP00 correspond to the scales $l$
which are much smaller than $\lambda$. Although
at these scales galactic rotational velocity is much larger than
the turbulent one, it's gradient is much smaller than the 
turbulent one. It is because of that the LP00
results in the presence of Galactic rotation and without it
coincide. The regime when the small
scales turbulent shear exceeds the regular one is natural.
Indeed, flows with high Reynolds number 
produce turbulence with shear larger that that of the original flow.
This is true, for instance,
for Couette flows in the presence of the moderately strong magnetic field
(Velikhov 1959, Chandrasechar 1962, Balbus \& Hawley 1991). The corresponding
criterion on the magnetic field is usually satisfied in the interstellar
medium.  

If gas is confined in an isolated cloud of size $S$ and the
galactic shear over this scale is neglected we get
\begin{equation}
\xi_s({\bf R},v) \sim
\int_{-S}^S {\mathrm d}z (1-\frac{|z|}{2S})
\; \frac{\xi({\bf r})}{[D_z({\bf r})+2\beta]^{1/2}}
\exp\left[-\frac{v^2}{2 (D_z({\bf r})+2\beta)}\right],
\label{ksicloud}
\end{equation}
where again somewhat ugly prefactor entering eq.~(\ref{AppB:2point_cloud})
is omitted.
Turbulent effects remain important up to the scale of the cloud $S$ at which
turbulent structure function saturates at the value $\sim (CS^m)^{1/2}$. 
The cloud size $S$ now plays the role of the
scale $\lambda$.
This observation allows to translate the LP00 results mostly written
for Galactic HI with rotation curve mapping to a case of an individual cloud.
Rigorous calculations
provided in LP00 prove that for scales much smaller than $S$ one can
use the results obtained for infinite medium in the presence of shear
and substitute $\lambda$ for $S$.  

When the amplitude of fluctuations grows with separation, one should use
structure functions in PPV. The transfer from one type of statistics to
another is similar to that discussed in Appendix~A.

Important feature of the PPV space is that $\rho_s({\bf X},v)$
exhibits fluctuations even if the flow is incompressible
and no density fluctuations are present.
Indeed, when one substitutes the expanded expression (\ref{Appeq:xi})
$\xi({\bf r}) =  \bar \rho^2 + \bar\rho^2(r_0/r)^\gamma$ into
eqs~(\ref{ksi2}-\ref{ksicloud}), both terms will give rise to non-trivial
contributions to $\tilde \xi_s({\bf R},v)$.
We shall therefore split the result correspondingly
\begin{eqnarray}
\tilde \xi_s({\bf R},v) &=& \tilde \xi_v({\bf R},v)
	+ \tilde\xi_\rho({\bf R},v), \nonumber \\
\tilde d_s ({\bf R},v) &=& \tilde d_v({\bf R},v)+ \tilde d_\rho ({\bf R},v). 
\end{eqnarray}
with the $v$-term describing pure velocity effects, while the $\rho$-term
arising from the actual real space density inhomogeneities that are
modified
by velocity mapping. To simplify the notation we have dropped index
$s$ from the right-hand-side quantities, since this split is only meaningful
in PPV space. In Appendix~C we discuss asymptotic  
small-R scalings of different contributions to PPV structure
and correlation functions.

\section{Turbulent Statistics and Radiative Transfer}

We start with the standard equation of radiative transfer (Spitzer 1978)
\be
dI_{\nu}=-g_{\nu} I_{\nu} ds+j_{\nu}ds~~~.
\label{transer1}
\ee
 In
the  case of self-adsorbing
emission in spectral lines that is proportional to first power of
density:
\be
g_{\nu}=\alpha(z) \rho(z) \phi_v(z)~~~,
\ee
\be
j_{\nu}=\epsilon \rho(z) \phi_v(z)~~~,
\ee
where $\phi_v(z)$ is given by (\ref{phi}).

A solution of this equation if no external illumination is present is
\be
I_v=\epsilon\int^S_0 dz \rho(z) \phi_v(z)\exp\left(-\int_{0}^z \alpha (z')
\rho(z')\phi_v(z')dz'\right)~~~.
\label{intensity}
\ee
To integrate 
(\ref{intensity}) we shall assume that $\alpha$ 
is constant. This constancy is the essence of the
Sobolev approximation that was found to be useful in many astrophysical
applications.  In this approximation we can use an integration variable
\be
Y_v(z)=\int_{0}^z \rho(z')\phi_v(z')dz'~~,
\label{Y}
\ee
which value at $x=S$ coincides with the density in PPV coordinates,
$ Y_v(S) = \rho_s({\bf X},v) $, and $Y_v(0)=0$, to integrate
eq.(\ref{intensity})
\be
I_v({\bf X})=\epsilon \int^{\rho_s}_0 dY_v {\mathrm e}^{-\alpha Y_v}=
\frac{\epsilon}{\alpha}\left[1-{\mathrm e}^{-\alpha \rho_s({\bf X},v)}\right]~~~.
\label{simplified}
\ee
In the case of vanishing absorption, the intensity is given by
the linear term in the expansion of the exponent in eq.~(\ref{simplified})
\be
I_v({\bf X})=\epsilon \rho_s({\bf X},v)~~.
\ee
and reflects the PPV density of the emitters.
If, however, the absorption is strong, the intensity of the emission
is saturated at the value $\epsilon/\alpha$ wherever 
$\rho_s({\bf X},v) \gg 1/\alpha$. Identification of the low contrast residual 
fluctuations may be difficult in practice.

The approximate expression given by eq. (\ref{simplified}) can be used
to find the statistics of the expected emissivity fluctuations. First of all,
the mean profile of the line is given by
\be
\langle I_v({\bf X}) \rangle =
\frac{\epsilon}{\alpha}\left[1-\langle{\mathrm e}^{-\alpha \rho_s({\bf X},v)}\rangle\right]~~~.
\label{meanprofile}
\ee

Our main goal is to calculate the structure function of the observed emissivity
\be
{\cal D}({\bf R})\equiv \left\langle \left(\int I_v({\bf X}_1)W_v dv -
\int I_v({\bf X}_2)W_v dv\right)^2\right\rangle,~~~
{\bf R} = {\bf X}_1 - {\bf X}_2 ~~~.
\label{dr_emiss}
\ee
Here the window function $W_v$
describes how the integration over velocities is
performed. When $W_v\equiv 1$ the
integration is being performed over the whole line as it is
a frequent case for CO turbulence
studies (see Falgarone et al. 1998, Stutzki et al. 1998)
while measurements in velocity slices of PPV data cube (channel maps) 
correspond to $W_v$ strongly peaked at
a particular velocity. Within the
VCA technique by  varying the width
of velocity channels one can obtain statistics of turbulent
velocities and density inhomogeneities of emitting medium.
Note, that 
the minimal width of the velocity channel is determined by the resolution
of an instrument.

Using elementary identities to transfer from averaging over 2D intensities
to averaging the underlying 3D fluctuations (see Lazarian 1995),
in the limit when the
absorption can be neglected we rewrite the expression (\ref{dr_emiss}) 
in the form
\begin{equation}
{\cal D}({\bf R}) = \epsilon^2 \int\, dv_1 W(v_1) \int\,dv_2 W(v_2)
\left[ d_s({\bf R},v_1,v_2) - d_s(0,v_1,v_2) \right]
\label{dzeroa}
\end{equation}
where 
homogeneity in physical directions is assumed. For infinite emitting medium
homogeneous turbulence produces a homogeneous image in the velocity space.
For finite emitting cloud the PPV image is approximately homogeneous over
velocity separations much less than the Doppler linewidth.  
Notably,  the combination
in brackets does not depend on the mean Doppler broadened 
profile of the spectral line,
$d_s({\bf R},v_1,v_2) - d_s(0,v_1,v_2)=\tilde d_s({\bf R},v_1,v_2) 
 - \tilde d_s(0,v_1,v_2)$.  Thus, it is sufficient to assume
that in the $v$-direction {\it fluctuations} of $\rho_s$ are
homogeneous. In this case, 
\be
{\cal D}({\bf R}) = \epsilon^2 \int ~dv W(v) \left[ d_s({\bf R},v) - d_s(0,v) \right]
\label{eq:a0limit}
\ee
with $W(v) = \int dv_+ W(v_+ - v) W(v_+ +v)$.
 The formula
(\ref{eq:a0limit}) together with Appendix B can be taken as a
starting point to reproduce the results in LP00 using structure functions
rather than spectra (compare with LP00). This
is demonstrated in Appendix D.

Substitution of the expression (\ref{simplified}) into 
(\ref{dr_emiss}) gives
\begin{eqnarray}
{\cal D}({\bf R})&=&\frac{\epsilon^2}{\alpha^2} 
\int\, dv_1 W(v_1) \int\,dv_2 W(v_2)
\times \nonumber \\
&\times& \langle
e^{-\alpha ( \rho_{11} + \rho_{12} )}
+ e^{-\alpha ( \rho_{22} + \rho_{21} )}
- e^{-\alpha ( \rho_{21} + \rho_{12} )}
- e^{-\alpha ( \rho_{11} + \rho_{22} )} 
\rangle
\label{28}
\end{eqnarray}
where we have used a shorthand notation $\rho({\bf X}_i,v_j) = \rho_{ij}$.
Homogeneity in spatial directions dictates that after averaging
the first term is equal
to the second one and the third is equal to the fourth
(e.g. for the density itself $\langle \rho_{11}+\rho_{12} \rangle
= \langle \rho_{22}+\rho_{21} \rangle $ and
$\langle \rho_{12}+\rho_{21} \rangle
= \langle \rho_{11}+\rho_{12} \rangle $).

We shall now derive an approximate expression for ${\cal D} ({\bf R})$,
applicable for small separations $R$.
Let us rewrite eq.~(\ref{28}) as 
\begin{eqnarray}
{\cal D}({\bf R})&=&\frac{\epsilon^2}{\alpha^2}
\int\, dv_1 W(v_1) \int\,dv_2 W(v_2)
\times \nonumber \\
&\times& \langle
e^{-\alpha ( \rho_{11} + \rho_{12} )} \left[
1 + e^{-\alpha ( \rho_{22} + \rho_{21} - \rho_{11} - \rho_{12} )}
- e^{-\alpha ( \rho_{21} - \rho_{11} )}
- e^{-\alpha ( \rho_{22} - \rho_{12} )} \right]
\rangle ~~.
\label{33}
\end{eqnarray}
It is easy to see that 
the term in brackets depends only on the differences in density taken 
between two lines of sight at the same velocity, i.e.
$\rho_{11}-\rho_{21}$ and $\rho_{22}-\rho_{12}$.
The point of rearranging the terms in this way is that for small separations
$R$ these differences are small, so we can retain only the leading order in
the power series expansion for the corresponding exponentials
\be
{\cal D} ({\bf R}) \sim \epsilon^2
\int\, dv_1 W(v_1) \int\,dv_2 W(v_2)
\; \langle
e^{-\alpha ( \rho_{11} + \rho_{12} )} \left[
\left(\rho_{11}-\rho_{21}\right)
\left(\rho_{12}-\rho_{22}\right)
\right]
\rangle ~~.
\label{DRexpansion}
\ee
The structure of the resulting expression is as follows. The term
in brackets is similar to the integrand in eq.~(\ref{dzeroa})
(indeed, $\langle \left(\rho_{11}-\rho_{21}\right)
\left(\rho_{12}-\rho_{22}\right) \rangle =  d_s({\bf R},v_1,v_2) - d_s(0,v_1,v_2) $
) while the effect of the absorption is manifested in the exponential
term which depends on the density distribution in velocity direction along
the line of sight. As we see, the series expansion we have performed at
small separation $R$ is indeed an expansion that assumes
$\alpha^2 \left[ d_s({\bf R},v_1,v_2) - d_s(0,v_1,v_2) \right]$ to be
small. Similar expansion 
is then applicable for any combination of structure functions that
vanishes at $R=0$ and this opens a way of dealing with non-Gaussian
statistics.

Although there is a cross-correlation between the two terms
in the integrand and statistical averaging cannot, in a general case, be factorized,
the correction for cross-correlation is of a higher order of smallness than
the leading term.  This can be seen by expanding the remaining exponential term
in power series. We can illustrate this statement by 
specifying Gaussian statistical distribution for $\rho_s$ and performing
explicit averaging which then gives
\begin{equation}
{\cal D} ({\bf R}) \sim \epsilon^2
\int\, dv_1 W(v_1) \int\,dv_2 W(v_2)
e^{\frac{1}{2} \alpha^2 \langle(\rho_{11}+\rho_{12})^2\rangle}
\left[d_s({\bf R},v_1,v_2) - d_s(0,v_1,v_2) + \alpha^2 d_{12} \right]
\label{31}
\end{equation}
The cross-correlation combination $d_{12}$ \footnote{$d_{12}=
\left[ d_s({\bf R},v_1,v_2) - d_s(0,v_1,v_2) +
d_s({\bf R},v_1,v_1) \right]
\left[ d_s({\bf R},v_1,v_2) - d_s(0,v_1,v_2) +
d_s({\bf R},v_2,v_2) \right]  $}
is vanishing as $R \to 0$
and $\alpha^2 d_{12}$ is small, as expected.
For non-Gaussian statistics the irreducible higher order correlations
appear. However it is suggestive that these correlations
provide vanishing contribution for $R\rightarrow 0$. Without providing
a formal derivation for the general case, we propose that
eq.~(\ref{DRexpansion}) can be rewritten as\footnote{A formal derivation
of this in the case of Gaussian fluctuations is given in Appendix~E. Our
efforts above were directed to show that the {\it structure} 
of the expression
does not depend on the assumption of the Gaussianity.}
\be
{\cal D}({\bf R}) \sim \epsilon^2
\int\, dv_1 W(v_1) \int\,dv_2 W(v_2)
\langle e^{-\alpha(\rho_{11}+\rho_{12})}\rangle
\left[d_s({\bf R},v_1,v_2) - d_s(0,v_1,v_2)\right] ~~.
\label{eq:dmaingen}
\ee 
or when approximation of homogeneity in velocity direction is warranted
\begin{equation}
{\cal D}({\bf R}) \sim \epsilon^2
\int\, dv_1 W(v_1) \int\,dv_2 W(v_2)
\langle e^{-\alpha(\rho_{11}+\rho_{12})}\rangle
\left[d_s({\bf R},v) - d_s(0,v)\right] ~.
\label{dmain}
\end{equation}

Expressions (\ref{eq:dmaingen}) and (\ref{dmain}) 
show that for sufficiently small $R$ the structure functions of intensity
differ from the earlier studied case by the window function determined
by the absorption
$W_{absorption}=\langle e^{-\alpha(\rho_{11}+\rho_{12})}\rangle$.
Two important conclusions directly follow from 
this observation. First, 
it is clear that if the window function determined
by data slicing is much narrower than that given by the  absorption
we get results indistinguishable from the earlier studied
case of no absorption. Second, in the case of integration over
the whole line of sight, the results will be different from the
earlier studied case because the window function is not equal
to unity any more.
The criterion when the absorption is not important for velocity
studies is straightforward. Expression (\ref{dmain}) transfers
into (\ref{eq:a0limit}) when $W_{absorption}$ is of the order
of unity over the range of scales studied.  

\section{Scaling Regimes of the Emissivity Statistics}

In this section we shall study emissivity statistics in the
regimes when it exhibits power-law scaling.
Such behavior is possible at the small scales where the
effect of the absorption is limited
$\alpha^2 \left[ d_s({\bf R},v_1,v_2) - d_s(0,v_1,v_2) \right] < 1$ 
and eq.~(\ref{dmain}) provides a good approximation.
We shall also restrict our consideration to the scales 
smaller than the cloud size $R/S \ll 1$ so that inhomogeneous
boundary effects can be neglected.

\subsection{Integrated Lines}

The presence in eq.~(\ref{dmain}) 
of the window function determined by absorption, i.e.
$\langle e^{-\alpha(\rho_{11}+\rho_{12})}\rangle$, changes
substantially results compared to the case of no absorption
discussed in LP00. Indeed, the integration over the whole spectral
line in the latter case removes the dependence of intensity
statistics on velocity. The window function determined by 
absorption defines to what extend the integration over
velocities is performed.

Let us estimate the form of the window assuming Gaussian statistics
and homogeneity of the density fluctuations
$\delta \rho_s = \rho_s - \langle \rho_s \rangle $. Then (see eq.~(\ref{e3}))
\begin{equation}
W_{absorption}=e^{-\alpha \langle \rho_s({\bf X_1},v_1) \rangle}
e^{-\alpha \langle \rho_s({\bf X_1},v_2) \rangle}
e^{\alpha^2 \langle \delta\rho_s^2({\bf X_1},v_1) \rangle}
e^{\alpha^2 \langle \delta\rho_s^2({\bf X_1},v_2) \rangle}
e^{-\frac{\alpha^2}{2}{\tilde d}_s(0,v)}~~~~.
\label{Wabs_gaussian}
\end{equation}
For homogeneous statistics the variance of the fluctuations does not 
depend on velocity thus the corresponding factors are constant and
affect only the normalization of the result. However, their presence reveals
limitations of the Gaussian approximation when the absorption is large.
Indeed, a similar term appears in the mean line profile
eq.~(\ref{meanprofile}) computed for Gaussian $\rho_s$, namely, 
$\langle I({\bf X},v_1)\rangle=
\frac{\epsilon}{\alpha}\left[1-\exp(-\alpha  \langle \rho_s({\bf X_1},v_1)
\rangle+ \frac{\alpha^2}{2}\langle \delta\rho_s^2({\bf X_1},v_1) \rangle)\right]$.
Clearly the answer is unphysical when 
$\alpha > 2 \langle \rho_s({\bf X_1},v_1)
\rangle/\langle \delta\rho_s^2({\bf X_1},v_1) \rangle$.
The problem arises because
the Gaussian fluctuations do not obey the constraint 
that the density $\rho_s$ is positive. Therefore for
high absorption, negative density excursions, however rare, dominate the 
result.
In view of this, we select in eq.~(\ref{Wabs_gaussian})
the factors which describe
the variable part of the mean intensity profile,
and write them in non-expanded form
\begin{equation}
W_{absorption}=\langle e^{-\alpha \rho_s({\bf X_1},v_1)} \rangle
\langle e^{-\alpha \rho_s({\bf X_1},v_2) } \rangle
e^{\frac{\alpha^2}{2} \langle \delta\rho_s^2({\bf X_1},v_1)+ 
\delta\rho_s^2({\bf X_1},v_2) \rangle}
e^{-\frac{\alpha^2}{2}{\tilde d}_s(0,v)}
\label{Wabs_gaussian_final}
\end{equation}
This approximate formula does not suffer from the defect we have mentioned.
The residual dependence on $\delta \rho_s$ variance can be traced to
the real effect of increase of intensity contrast
between two points if absorption is present. However, this factor is constant
in our treatment and we shall normalize it out.
We summarize our considerations in the formula
\begin{equation}
{\cal D}({\bf R}) \propto
\int\, dv \; \tilde W(v)\; e^{-\frac{\alpha^2}{2}{\tilde d}_s(0,v)} \;
\left[d_s({\bf R},v) - d_s(0,v)\right] ~,
\label{dmainapprox}
\end{equation}
where 
\begin{equation}
\tilde W(v) \equiv \int {\mathrm d} v_+ W(v_1) W(v_2)
\langle e^{-\alpha \rho_s({\bf X_1},v_1)} \rangle
\langle e^{-\alpha \rho_s({\bf X_1},v_2)} \rangle~~~~~.
\end{equation}
The most important effect induced by absorption is an additional exponential
down-weighting in the projection of the contribution from the points
with large velocity separation $v$ in a manner which
itself depends on the turbulence statistics. For subsonic turbulence
where the velocity variance is dominated by the thermal dispersion and
is not scale dependent, the effect of absorption in our approximation
amounts to a constant and for statistics is equivalent to zero absorption and
zero turbulent velocity case. However, one should still check
whether the linearization of the kernel holds at the scales under
study for a given absorption value.

The most important qualitative
characteristic of the window is its width, which for absorption we shall
define as velocity $v_{ab}$ at which 
\begin{equation}
\alpha^2 \tilde d_s(0,v_{ab}) =1~~~~~~~.
\label{vab}
\end{equation}
In terms of the velocity-density decomposition of the PPV structure function,
the product of the two windows arise $\sim e^{-\alpha^2 \tilde d_v(0,v)/2}
e^{-\alpha^2\tilde d_\rho(0,v)/2}$. Both factors act simultaneously 
but the one with the smallest width determines the gross effect.
Asymptotic analysis of $\tilde d_s(0,v)$ in Appendix C gives for the
absorption window widths
\begin{eqnarray}
v_{ab}/D_z(S)^{1/2}
 &\approx& \left(\alpha \bar \rho\right)  ^{\frac{2m}{m-2(1-\gamma)}}, 
~~~~~~ m > 2/3 \; (1-\gamma) \label{eq:abs_width1}\\
v_{ab}/D_z(S)^{1/2}
 &\approx& \left(\alpha \bar \rho\right)^{-1} ~~~~~~~,
~~~~~~ m < 2/3 \; (1-\gamma)
\label{eq:abs_width2}
\end{eqnarray}
where we have omitted numerical coefficients of order unity and estimated
the mean density in PPV space as
\begin{equation}
\langle \rho_s \rangle = \bar \rho S/D_z(S)^{1/2} ~~.
\end{equation}
 Absorption effects become negligible 
if $v_{ab}/D_z(S)^{1/2} \gg 1$.

{\it Effect of Velocity Fluctuations}\\
Let us consider first the $\delta\rho_s$ 
fluctuations that arise from velocity fluctuations
only. Indeed, even if the underlying density is constant, random
velocities do produce caustics that were shown in LP00 to be
very important for the analysis.

\begin{figure}[h]
\plotone{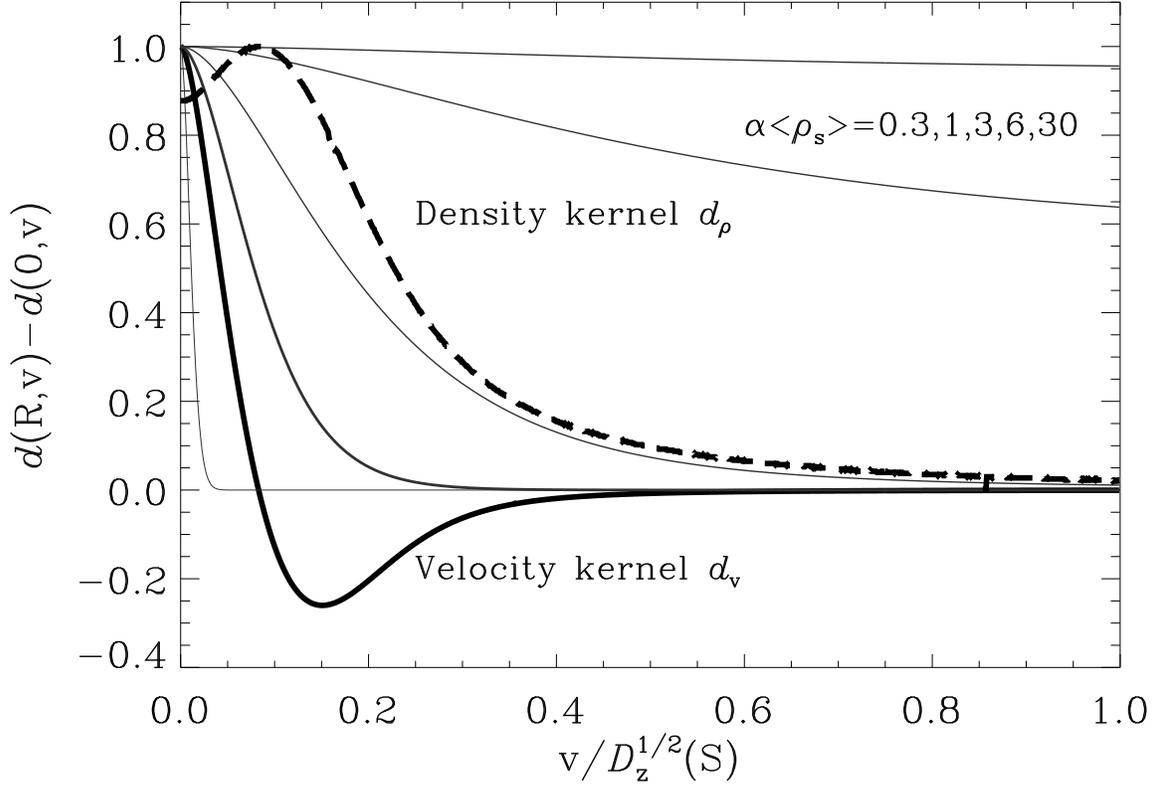}
\caption{The velocity kernel $d_v(R,v)-d_v(0,v)$ (thick solid line)
and the density kernel $d_{\rho}(R,v)-d_{\rho}(0,v)$ (dotted line) for 
$R/S=0.01$ and $m=2/3$ as a function of the line of sight velocity separation
$v/D_z^{1/2}(S)$ normalized over velocity
dispersion. The thin solid lines correspond to  the window function 
$\exp(-\alpha^2 {\tilde d}_s(0,v)/2)$ and are drawn for the values
$\alpha\langle \rho_s\rangle$ from $0.3$ to $30$ (from top to bottom).
For the displayed value of $R$, the linearized treatment of the kernels is
applicable up to $\alpha\langle \rho_s \rangle \sim \sqrt{2} (R/S)^{(m-2)/4}
\approx 6$, the corresponding line is highlighted.
}
\label{fig1}
\end{figure}

In Figure~\ref{fig1} we plot the 
correspondent kernel
$\left[d_v({\bf R},v) - d_v(0,v)\right] $ as a function of $v$
for a fixed sample value of $R$ and, simultaneously, 
the window function  $ e^{-\frac{\alpha^2}{2}{\tilde d}_s(0,v)} $ for the
range of absorption amount characterized by $\alpha \bar\rho_s$.
The kernel is highly peaked at the
zero velocity separation $v=0$, has region of negative values, and
approaches zero at large $v$. The window is unity for zero absorption 
but also peaks at $v=0$ as absorption increases.

Integration of the velocity kernel over the whole line in the absence of 
absorption and $W(v)=1$ gives zero, in concordance with the 
expectation of canceling all the velocity effects. 
In the presence of high absorption one might hope that 
down-weighting all nonzero $v$ contribution will act as an effective
{\it thin} slice reproducing {\it thin} slice power-law asymptotics
of $D(R)$, which is achieved when
the width of the slice becomes small compared to {\it rms} velocity at the
scale of the study $R$ (LP00). However we are restricted by the fact
that our
linearized approximation (\ref{DRexpansion}) is valid only when absorption
is moderate $\alpha^2 d_s({\bf R},0) < 1 $, while higher absorption (or larger
scales at fixed $\alpha$) will induce nonlinear modifications of the result.
Thus we find {\it thin} slice scaling is never realized for $m \ge 2/3$.
Instead, the new universal
intermediate power-law scaling is exhibited for moderate absorption.
Dependence of the 2D structure function on the 
underlying velocity index $m$ is lost in this regime
For $m < 2/3$ there is a range of scales when absorption can modify 
the scaling to that of the {\it thin} slice in which case sensitivity to $m$
is present.

Analytical estimates can provide some insight into the expected behavior
of ${\cal D}_v({\bf R})$ after integration as $R \to 0$.
The value of the kernel at the peak, $d_v({\bf R},0)$ is given by
the asymptotic expression (\ref{AppC:eq_dvR})
(see also LP00 for asymptotics
of the correlation functions in the velocity space),
\begin{equation}
d_v(R,0) \sim \left|
\frac{\sqrt{\pi}\;\Gamma(m/4-1/2)}{\Gamma(m/4)}\right| R^{1-m/2}~~~,
\label{DR0estim}
\end{equation}
where we are more interested in the $R$ dependence of the expression
than in the numerical prefactor.
Incidently, this gives a criterion on the validity of 
the linear expansion of the velocity kernel.
With dimensional quantities restored the condition
\begin{equation}
\alpha^2 \langle \rho_s \rangle^2 < 
\left| \frac{\Gamma(m/4)}{\sqrt{\pi} \Gamma(m/4-1/2)} \right|
\left( \frac{S}{R} \right)^{1-m/2}
\label{eq:alphalinear}
\end{equation}
sets the scale range where our treatment is applicable.

Let us first consider the case $m \ge 2/3$. 
With the width of absorption induced window given by eq.~(\ref{eq:abs_width1}),
the condition (\ref{eq:alphalinear}) requires
 $v_{ab}^2 \gtrsim D_z(S) (R/S)^m $,
i.e velocity cutoff exceeds rms turbulent velocity at scale $R$ and thus
is insufficient to reproduce {\it thin} slice which require opposite relation
 to be true (see LP00).

This is, however, sufficient to suppress the negative
large $v$ tail in which case 
the dominant contribution to the integral  scales as the height of the kernel
peak times its width.
The scaling of the width of the kernel peak with $R$ 
can be readily obtained from the following consideration:
If we integrate the kernel over $v$ choosing fixed integration range
$dV$ but varying $R$ from large values to the small, then
the projected result will scale with $R$ according to 
{\it thin} slice asymptotics while $D_z(R)^{1/2}> dV$.
The change to the {\it thick} slice behavior takes place
when the condition is reversed.
But formally in the {\it thin} slice regime the integral is determined by the
zero-lag peak behavior only, while in the {\it thick} slice integration 
encompasses most of the kernel. Therefore $dV = D_z(R)^{1/2} $, where $D_z(R)$
is the $z$ projection of the correlation tensor of velocity (see eq.(\ref{eq:Dz}),
represents the required estimate of the width of the kernel. This scaling
is valid both for the zero crossing point of the velocity kernel and
the width of the density kernel that we consider later.

The total scaling of the 2D emissivity structure function is thus
${\cal D}_v(R) \sim d_v(R,0) dV \propto R^{1-m/2} R^{m/2} \propto R$,
with dependence on $m$ canceled out.
The corresponding 2D spectra scales as $K^{-3}$.

Numerical integration using  eq.~(\ref{31}) confirms
that indeed the velocity statistics exhibit
a new intermediate asymptotic regime, which is characterized by a universal
spectral index of $-3$. In Fig.~2 we show structure
functions of intensity for two different indexes $m=2/3$ and
$m=1$. The range for the universal asymptotics is clearly seen.

\begin{figure}[ht]
\plottwo{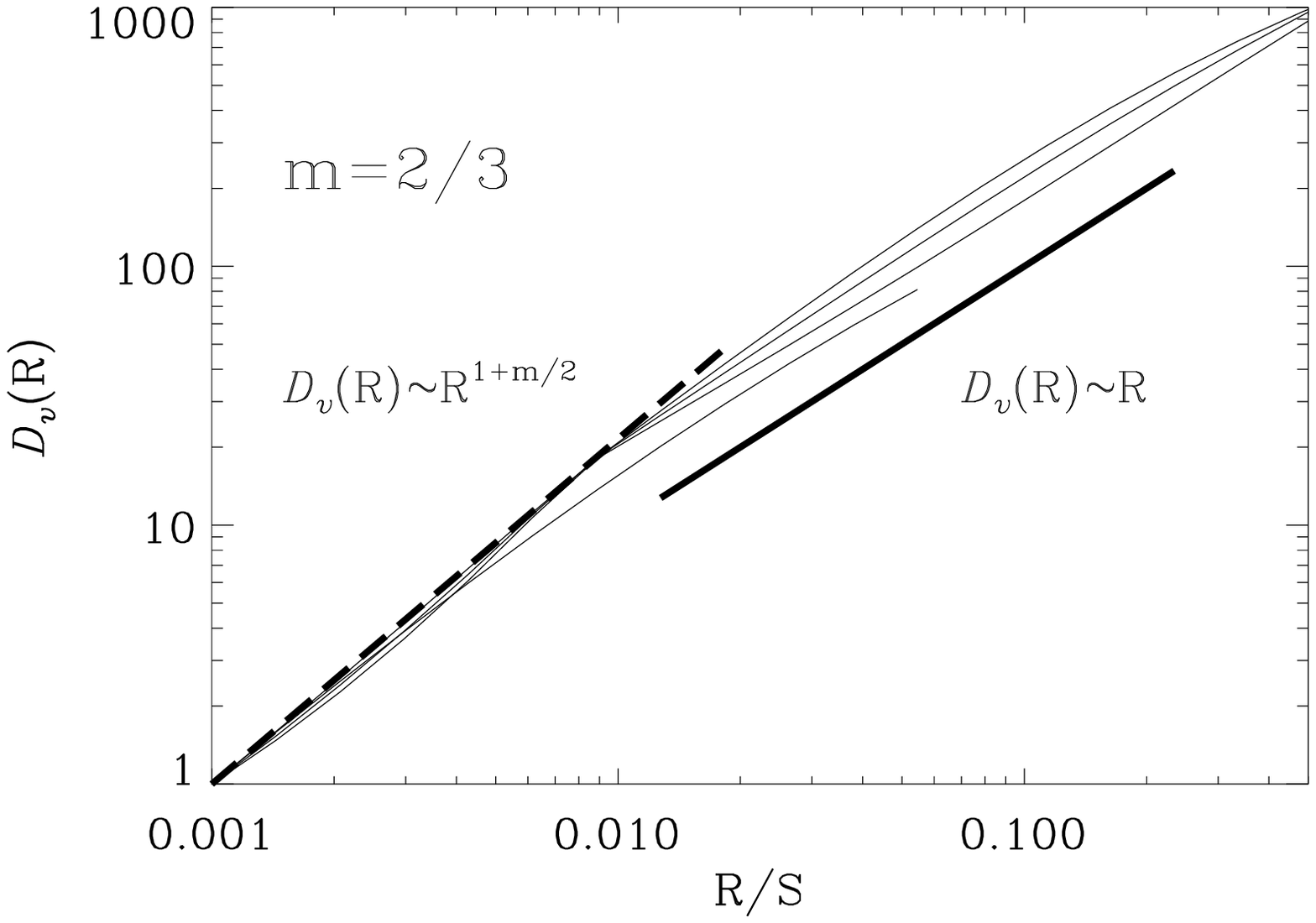}{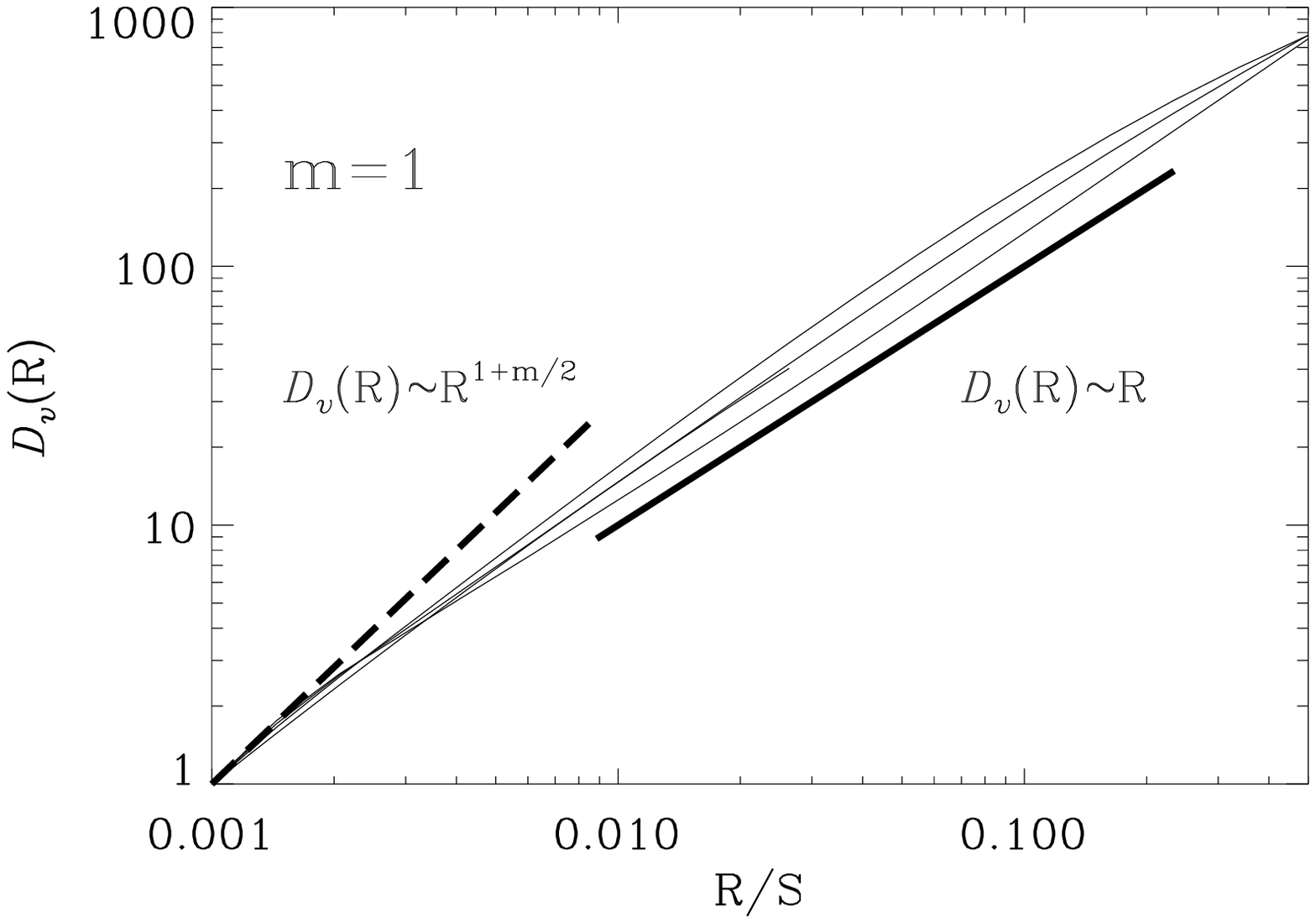}
\caption{ The structure function of intensity for $m=2/3$ (right
panel) and $m=1$ (left panel) for a set of absorption
coefficients $\alpha \langle \rho_s \rangle = 0.2,0.36,0.7,2.5 $
(from top to bottom).
We use the general non-linear expressions from Appendix~E, but 
plot the curves only up to $R$ when linearized theory is still accurate.
At the intermediate scales $R$
the structure function tends to the universal regime
$\sim R$ which does not depend on the particular index $m$.
The upper dashed line corresponds to the {\it thick} slice asymptotics
for the velocity (spectral index $-3-m/2$) and the lower
dotted line correspond to the universal index $-3$ that
does not depend on $m$.
}
\label{fig2}
\end{figure}
Fig.~2 also shows that for very small $R$ the structure function
grows faster than $R$, which entails
spectra steeper than $K^{-3}$.
Indeed, for sufficiently small $R$ the $\exp(\alpha^2 {\tilde d}_v(0,v))$ window
does not cut out even the distant region where the kernel gets negative
in which case we expect to see the thick slice asymptotics
$\sim R^{1+m/2}$ (i.e $K^{-3-m/2}$ for the spectrum, see LP00)
in accordance with our numerical calculations.
Thermal effects (not included in numerical experiment)
at scales where turbulence is subsonic $D_z(R) < \beta$
lead to even faster fall-off of the structure function at small $R$ and
indeed make any coherent scaling of turbulent velocity hard to observe
in emissivity statistics at that scales.

For $m < 2/3$, on the other hand, absorption may lead to {\it thin} slice
scaling $\propto R^{1-m/2}$ for some intermediate range of scales before
$R$ is so large as to cause nonlinear deviations. Indeed, in this case
$\left(\alpha \bar \rho\right)\approx v_{ab}/D_z(S)^{1/2}$ and condition
(\ref{eq:alphalinear}) can be rearranged to read 
$ D_z(R)/v_{ab}^2 (R/S)^{1-3m/2} < 1 $ which can be satisfied even
for relatively high absorption $ v_{ab}^2 /D_z(R) > 1$ correspondent to 
{\it thin} slice velocity cutoff. For a given absorption parametrized by
$v_{ab}$ we summarize the behaviour at the different scales
\begin{table}[t]
\begin{displaymath}
\begin{array}{lll}
\mathrm{Scale~Range} & \mathrm{Intensity~Scaling} & \mathrm{Regime}\\
\hline \\
R/S < (\beta/D_z(S))^{1/m} & \mathrm{velocity~effects~erased} & 
(\mathrm{subsonic~regime}) \\
R/S \ll \left(\frac{v_{ab}^2}{D_z(S)}\right)^{1/m} &
{\cal D}(R) \propto R^{1+m/2} &  (thick~\mathrm{slice})
\label{eq:DVR_thick} \\
R/S < \left(\frac{v_{ab}^2}{D_z(S)}\right)^{1/m} &
{\cal D}(R) \propto R^1 & (\mathrm{intermediate~scaling})
\label{eq:DVR_interm} \\
\left(\frac{v_{ab}^2}{D_z(S)}\right)^{1/m} <
R/S < \left(\frac{v_{ab}^2}{D_z(S)}\right)^{2/(2-m)} &
{\cal D}(R) \propto R^{1-m/2} & (thin~\mathrm{slice})
\label{eq:DVR_thin} \\
\left(\frac{v_{ab}^2}{D_z(S)}\right)^{2/(2-m)} < 
R/S & \mathrm{not~ a~ power ~law}
& (\mathrm{strong~absorption~regime})
\label{eq:DVR_nonlinear}
\end{array}
\end{displaymath}
\caption{Scalings of structure functions of intensity fluctuations arising
from velocity fluctuations for the power-law
underlying 3D velocity statistics. In the strong absorption regime ${\cal D}$
does not follow a simple power-law. {\it Thin} slice regime
does not exist for $m>2/3$. $D_z(S)$ is given by eq.(\ref{eq:Dz}).}
\label{table:scaling_regimes}
\end{table}
in the Table~\ref{table:scaling_regimes}.
All four scaling regimes are present for $m < 2/3$ although how pronounced is
actually {\it thin} slice scaling (\ref{eq:DVR_thin}) depends on 
the extend of the scale range $(v_{ab}^2/D_z(S))^{1/m}$ ---
$(v_{ab}^2/D_z(S))^{2/2-m}$, which grows with absorption strength if $m < 2/3$.
This interval does not exist for $m \ge 2/3$.
We note that the two-dimensional emissivity statistics of integrated
lines ${\cal D}_v(R)$ restore $m$ sensitivity for $m < 2/3$ where
the line-of-sight statistics $\tilde d_v(0,v)$
is saturated at $\propto v^2$. The latter behavior is not surprising. Indeed,
structure functions $\tilde d_v(0,v)$ cannot grow faster than $v^2$ and this
explains the saturation. However, the information along the line of sight is
not lost. The analysis of the spectra of fluctuations along the velocity
coordinate performed in LP00 confirms this.

{\it Effect of Density Fluctuations}\\
Density fluctuations modify the statistics of emissivity. 
Mathematically the density fluctuations are imprinted
on the $\xi_{\rho}$ part of the 3D correlation function.
This correlation function depends both on the statistics of density
and velocity. As we discussed earlier, density correlations may
be short-wave or long-wave dominated. Within power-law model for over-density 
statistics they are described,
respectively, by correlation $\tilde \xi(r) = \bar \rho^2 (r_0/r)^{\gamma}$
or structure $\tilde d(r) = \bar \rho^2 (r_0/r)^{\gamma}$ functions,
where $\gamma$
is positive in the former and negative in the latter case (see Appendix~A).
The amplitude of the density effect is encoded
in the correlation scale $r_0$ which has a meaning of the scale at
which density inhomogeneities have an amplitude equal the mean density
of the gas (i.e. $\delta \rho \sim \bar \rho $ for $\gamma > 0$ or
$\rho_1-\rho_2 \sim \bar \rho$ for $\gamma < 0$).
This gives rise to the amplitude factor $(r_0/S)^{\gamma}$ in the asymptotic
expressions of Appendix~C (e.g. eq.~(\ref{AppC:eq_drhoR})).

Contribution from density to PPV statistics combines linearly with the pure
velocity term and an important question of which term dominates arises.
Note, that
both terms contribute to the absorption window at the same time 
in non-linear fashion.

Table~2 summarizes at which scales the density inhomogeneity
contribution dominates velocity effects in PPV space.
\begin{table}[t]
\begin{displaymath}
\begin{array}{lllc}
\mathrm{Condition} & \mathrm{Case}~\gamma < 0
 & \mathrm{Case}~\gamma > 0 & \mathrm{Eqns} \\
\hline\\
\tilde d_{\rho}(R,0) > \tilde d_v(R,0): & R > r_0 &  R < r_0 &
(\ref{AppC:eq_dvR},\ref{AppC:eq_drhoR}) \\ \\
\tilde d_{\rho}(0,v) > \tilde d_v(0,v): & & & \\
m \ge \mathrm{max}\left[\frac{2}{3},\frac{2}{3}(1-\gamma)\right]
 & v^2 > D_z(S) (r_0/S)^m & v^2 <  D_z(S) (r_0/S)^m &
(\ref{AppCeq:d_vVfg23},\ref{AppCeq:d_rhoVf1})\\
\frac{2}{3} < m < \frac{2}{3}(1-\gamma)
& v^2 > D_z(S) (r_0/S)^{-\frac{2/3 \gamma m}{m-2/3}} &\mathrm{not~applicable}&
(\ref{AppCeq:d_vVfg23},\ref{AppCeq:d_rhoVf2})\\
\frac{2}{3}(1-\gamma) < m < \frac{2}{3}
 & \mathrm{not~applicable} & v^2 < D_z(S) (r_0/S)^{\frac{2/3 \gamma m}{m-2/3(1-\gamma)}} &
(\ref{AppCeq:d_vVfl23},\ref{AppCeq:d_rhoVf1})\\
m \le \mathrm{min}\left[\frac{2}{3},\frac{2}{3}(1-\gamma)\right]
& r_0/S < 1 & r_0/S > 1 &
(\ref{AppCeq:d_vVfl23},\ref{AppCeq:d_rhoVf2})
\end{array}
\end{displaymath}
\caption{Range of the scales where the impact of
density inhomogeneities to the PPV statistics exceeds
the velocity contribution. For steep spectra $\gamma < 0$,
we need $r_0 < S$ for density effects to be seen. This is
feasible in a model where high amplitude density inhomogeneities
are saturated at scales $r_c < S$.
For the validity of the asymptotic scaling used, 
$R \ll S~ (\mathrm{or}~r_c) $ is required. When $\tilde d_{\rho}(0,v) > \tilde d_v(0,v)$
the window function is determined by $\tilde d_{\rho}(0,v)$ (see eq.~(\ref{vab})). In
addition, the window function is determined by density fluctuations for subsonic velocities.
}
\label{table:density}
\end{table}
For steep spectra,
the density contribution dominates large scales, above characteristic
scale determined by $r_0$, For shallow spectra it is dominant
at small scales.
Density inhomogeneities also dominate observed statistics at  
small scales where velocity is subsonic.

What are the reasonable values for $r_0$ ?  We argue that the amplitude of
density perturbations of the scale of the cloud itself should not
significantly exceed the mean density, otherwise the applicability 
of the notion of a cloud is suspect. For a shallow spectrum $\gamma > 0$ this
means $r_0 < S$. For a steep spectrum $\gamma < 0$, $r_0 > S$ is likely,
if a power-law distribution of fluctuation amplitudes extends all the way
to the cloud scale $S$. In both cases $(r_0/S)^{\gamma} < 1$. 

An immediate conclusion from Table~\ref{table:density}
is that for steep spectra $\gamma < 0$
density terms are expected
to always be subdominant, since within a cloud $R \le S < r_0$.
However we must stress that this relies on an assumption that
the turbulent density scaling extends all the way to the cloud size $S$,
in which case high value of $d(S)$ will indicate large inhomogeneities
of the size of the whole cloud.
But if the density structure function saturates at smaller separation
$r_c < S$, {\it e.g.}
as $\tilde d(r)=\tilde d(\infty) r^{-\gamma}/(r^{-\gamma}+r_c^{-\gamma})$,
we may encounter a high amplitude 
$\tilde d(\infty) \gg \bar \rho^2$ without density at the scale of the cloud
being significantly perturbed, since all inhomogeneities are restricted
to smaller scales. As seen from the
expansion $\tilde d(r) \sim 
\tilde d(\infty) (r/r_c)^{-\gamma},~~ r < r_c$, the effective correlation scale 
in this case is $r_0=r_c \times [\bar\rho^2/\tilde d(\infty)]^{-1/\gamma} < r_c$
 and in the range of
scales $r_0 < R < r_c$ one will observe the density effects.
The physical picture, leading to such situation, is to have
strong energy injection at scale smaller than a cloud size,
with turbulent cascade establishing steep spectrum to smaller scales,
while having energy sources only weakly correlated at larger scales.

The Table~\ref{table:density} 
shows that for a shallow spectrum the density term is important for PPV
statistical descriptors in both $R$ and $v$ direction at small scales
$R < r_0$ and  $v^2 < D_z(r_0)$ (for $m \ge 2/3$).

The marked effect of absorption on intensity
fluctuations in integrated lines is that the density
contribution cannot always be recovered by increasing the slice thickness,
as it is the case of no or small absorption.
On the contrary, we are able to observe
the density impact on the emissivity only in restricted scale range.
The reason for this is that for a wide range of scales
above the thermal dispersion scale the velocity term is not suppressed in the
presence of the absorption even when we integrate over the whole line.

We shall now discuss the how the density contribution scales,
depending on separation $R$ and conditions on $\gamma$ and $m$.
Whether density term will dominate the overall PPV statistics depends
on the amplitude of density inhomogeneities given by $r_0$ according to
Table~\ref{table:density}.  Our results are summarized below in
Table~\ref{table:density_scaling}.

Scaling estimations for density are obtained in a similar way 
to the case of pure velocity term in the previous section.
The main ingredients are density kernel
and absorption windows, exemplified in Figure~\ref{fig1}.
The complication is that the absorption
window is determined by $d_s(0,v)$,
with contributions from both velocity and density giving
 rise to two multiplicative windows (see equation~(\ref{vab})).
The one with the smallest width of the two is the most important one.
We shall explicitly note when the details of the absorption window 
are important.

The peak of the density kernel now scales as $R^{1-\gamma-m/2}$ (see 
equation~\ref{AppC:eq_drhoR}) while its width is $dV \sim D_z(R)^{1/2}$, 
as we discussed before. Integrating over the whole line in the absence of
absorption gives an estimate 
${\cal D}(R) \propto R^{1-\gamma-m/2}R^{m/2}\sim R^{1-\gamma}$
(this refers to a correlation,
rather than to a structure function of emissivity if $\gamma > 1$).
which is exactly the {\it thick} slice regime of LP00. If absorption is not
too strong, $v_{ab} > D_z(R)^{1/2}$, it will not affect this result.
This is different from pure velocity term, since the velocity term exactly
vanishes when integrated over the whole velocity range in the absence of
absorption, and small absorption leads to qualitatively new 
intermediate "universal" scaling.

The condition of weak absorption
in the presence of density, that replaces the one in (\ref{eq:alphalinear}),
is given by
\begin{equation}
\alpha^2 \langle \rho_s \rangle^2 < 
\left| \frac{\Gamma(\gamma/2+m/4)}{\sqrt{\pi} \Gamma(\gamma/2+m/4-1/2)} \right|
\left( \frac{S}{R} \right)^{1-\gamma-m/2}
\label{eq:alphalinear_density}
\end{equation}
Using (\ref{eq:abs_width1},\ref{eq:abs_width2}) to express absorption via
$v_{ab}$, we find familiar constraints when absorption
effects are not strongly non-linear
\begin{eqnarray}
v_{ab}^2 &>& D_z(S) (R/S)^{m} \sim D_z(R), 
~~~~~~~~~~~~~~~~~~ m > 2/3 \; (1-\gamma) \label{eq:abs_width_linear1}\\
v_{ab}^2
 &>& D_z(S) (R/S)^{1-\gamma-m/2} ~~~~~,
~~~~~~~~~~~~~~~~~~ m < 2/3 \; (1-\gamma)
\label{eq:abs_width_linear2}
\end{eqnarray}

\begin{table}[t]
\begin{displaymath}
\begin{array}{lll}
\mathrm{Scale~Range} & \mathrm{Intensity~Scaling} & \mathrm{Regime}\\
\hline \\
R/S < (\beta/D_z(S))^{1/m} & {\cal D}(R) \propto R^{1-\gamma}  & 
(\mathrm{subsonic}) \\
R/S < \left(\frac{v_{ab}^2}{D_z(S)}\right)^{1/m} &
{\cal D}(R) \propto R^{1-\gamma} & (\mathrm{thick~\mathrm{slice}})
\label{eq:DrhoR_interm} \\
\left(\frac{v_{ab}^2}{D_z(S)}\right)^{1/m} <
R/S < \left(\frac{v_{ab}^2}{D_z(S)}\right)^{\frac{2}{2-2\gamma-m}} &
{\cal D}(R) \propto R^{1-\gamma-m/2} & (thin~\mathrm{slice})
\label{eq:DrhoR_thin} \\
\left(\frac{v_{ab}^2}{D_z(S)}\right)^{2/(2-m)} < 
R/S & \mathrm{not~ a~ power ~law}
& (\mathrm{strong~absorption})
\label{eq:DrhoR_nonlinear}
\end{array}
\end{displaymath}
\caption{Scalings of structure functions of intensity fluctuations arising
from density inhomogeneities for the power-law
underlying 3D velocity statistics. 
{\it Thin} slice behaviour require $\gamma < 1-3m/2$.
In the strong absorption regime ${\cal D}$
does not follow a simple power-law.}
\label{table:density_scaling}
\end{table}

The first case, $m > 2/3(1-\gamma)$ is applicable
for the most interesting values of $m$ if the density spectrum is shallow,
$\gamma > 0$. Here, linearity and {\it thick} slice conditions coincide and
one cannot achieve {\it thin} slice regime before the absorption effects
become nonlinear.

The second case, $m < 2/3(1-\gamma)$ has wide range of applicability for
steep spectra $\gamma < 0$. Here linearity condition is less stringent
and in the range of scales $(v_{ab}^2/D_z(r_c))^{1/m} < R/r_c < 
(v_{ab}^2/D_z(r_c))^{1/1-\gamma-m/2}$ absorption effects
modify the emissivity scaling towards 
the {\it thin} slope $R^{1-\gamma-m/2}$, while nonlinear effects come into 
play only at larger scales. 
\footnote{To be consistent, velocity effects of
Table~\ref{table:scaling_regimes}
have to be reevaluated in this
case due to larger contribution from density perturbations to the
absorption.}.

To summarize: For shallow spectra one expects line-integrated statistics for
density perturbations to follow
{\it thick} slice scaling until absorptions effects become too strong.
Density terms are visible in the overall emissivity statistics at
$R < r_0$.
For steep spectra, density contribution may exhibit, in addition to
{\it thick} slice behaviour at small scales, the {\it thin} slice scaling
at intermediate scales. However, density contribution can be dominant in
emissivity statistics only when $R > r_0$, which require high amplitude
density inhomogeneities saturated at scales smaller than a cloud size.

Table~3 and Fig.~3 illustrate how the statistics of emissivity
changes with the change of the
spectral index of density and velocity.

\begin{figure}[ht]
\plottwo{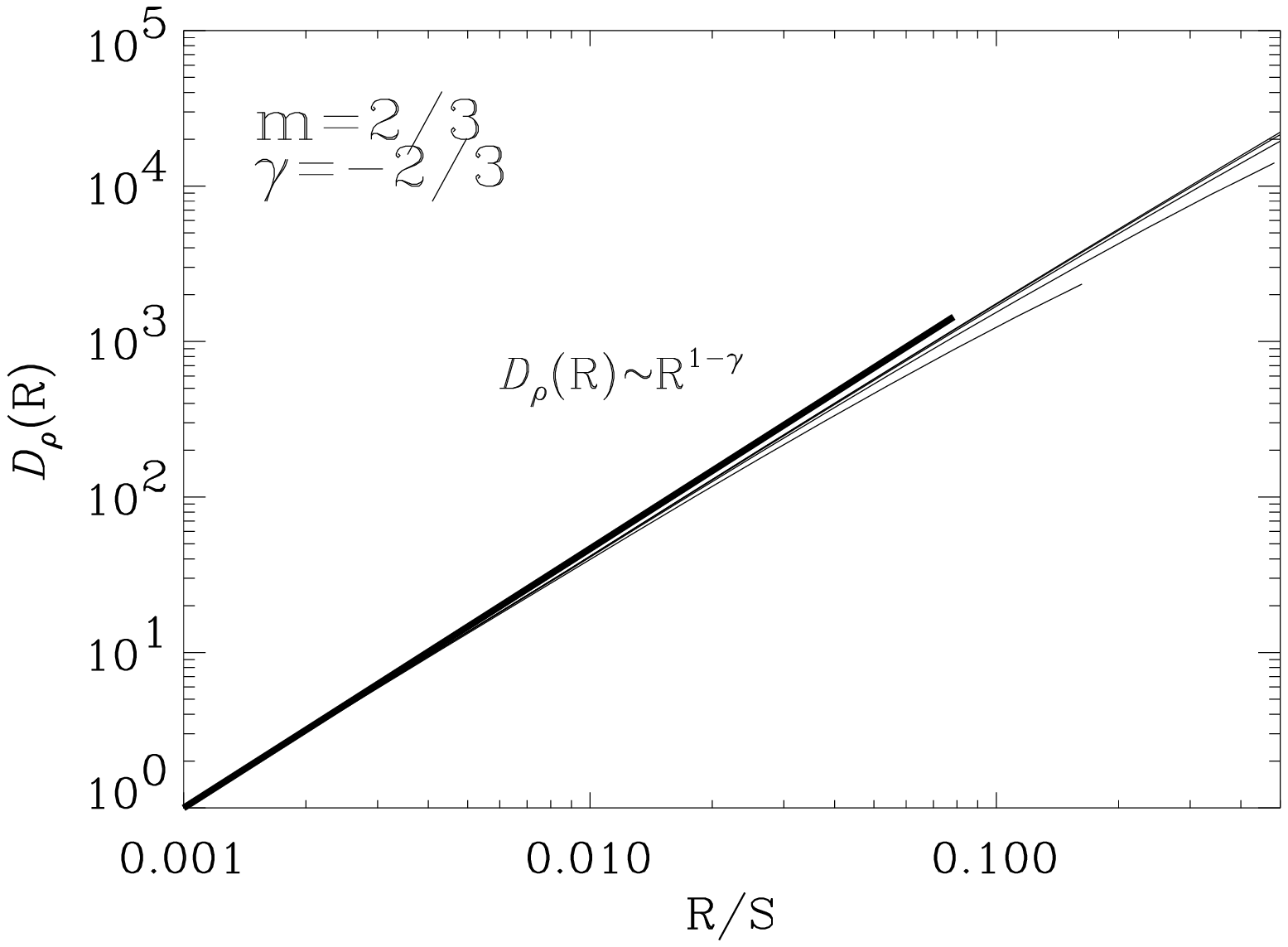}{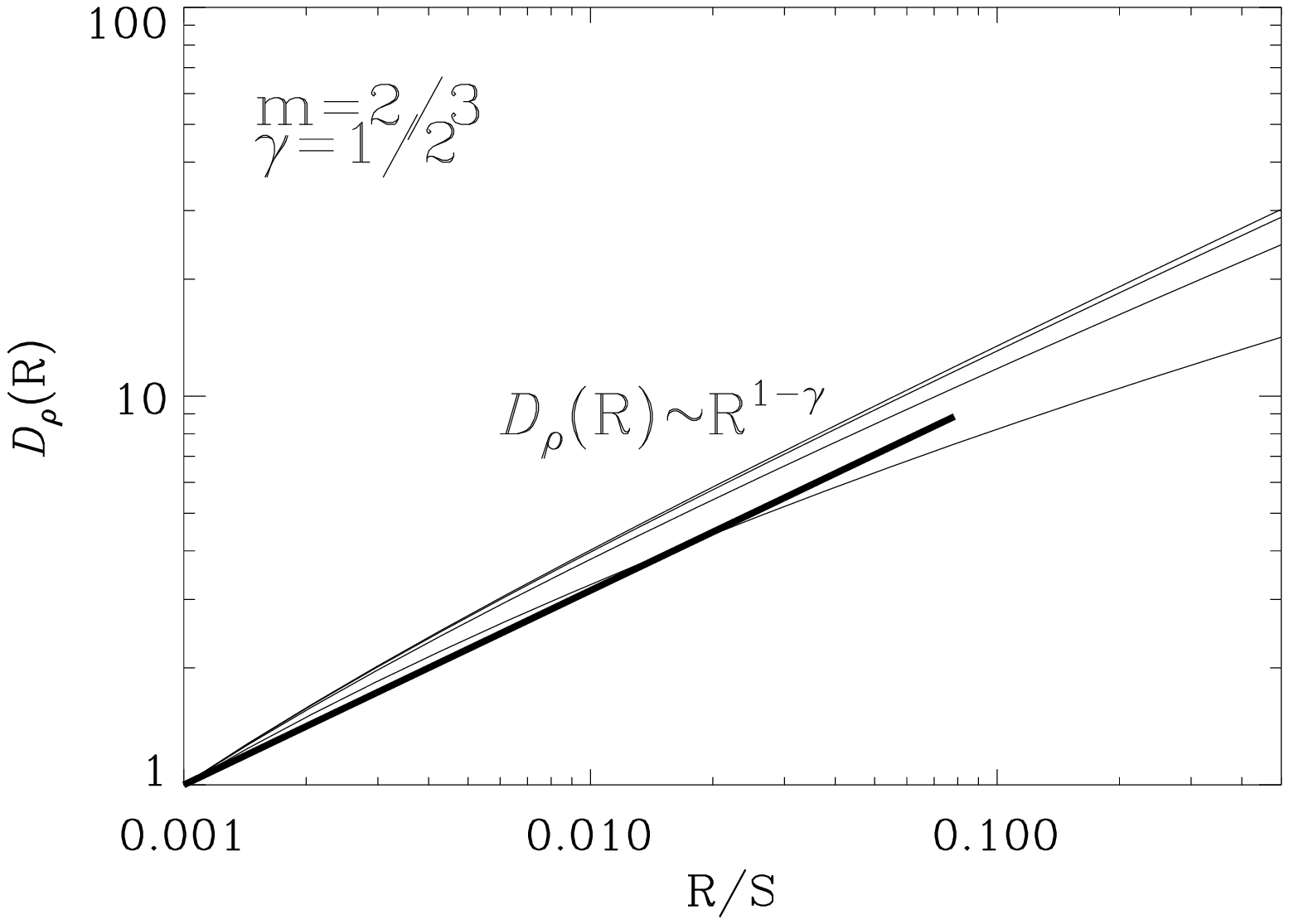}
\caption{Left panel: 
The structure function of intensity for Kolmogorov
density  and Kolmogorov velocity.
Right panel: The structure
function of intensity for shallow density $\gamma=1/2$
Kolmogorov velocity $m=2/3$. 
Absorption parameter for the lines shown varies in the range
$\alpha \langle \rho_s \rangle (r_0/S)^{\gamma/2} = 0.1$
(top) to $ 1 $ (bottom). Scaling solutions for both cases
are given in heavier line and labeled.
The calculations are done using unexpanded formula given by eq.~(\ref{largeD1}).
The non-linear effects get important as $R$ increases.}
\label{fig3}
\end{figure}

Can we ever observe density fluctuations in the absorbing gas
outside the ranges set by correlation scales, for example when
turbulent velocities are small ? 
While one decreases the amplitude of supersonic turbulence
the absorption effects increase due to atoms concentrating at
similar velocities which is expressed by PPV density 
$\bar \rho_s \approx \bar \rho S/\sqrt{D_z(S)+2\beta}$ growth, making
recovering turbulence scaling more difficult.
However velocity effects disappear as
turbulence becomes subsonic at all scales $D_z(S) \ll \beta$.
The power law index of density
inhomogeneities can then be obtained over the range of scales
if $\alpha \bar \rho S/\sqrt{\beta} \ll 1$.
Under similar conditions we can see density fluctuations at small
scales $r<r_{\rm thermal}$, where the turbulence is subsonic even if it is supersonic at
large scales. The scale $r_{\rm thermal}$ can be determined as
$D_z(r_{\rm thermal})=\beta$.

It may look somewhat counterintuitive that even in the absence of
velocity we never see a thin skin of the cloud exhibiting 2D {\it spatial}\footnote{We may note that a thin velocity slice is very different from
a thin slice in space. Integration over the entire spatial extend of
the cloud is
performed while the velocity slice is defined.} 
slice of density fluctuations. As absorption is proportional to density
in our model, the optical depth changes with the change of density. As the
result the emission {\it fluctuations} are not dominated
by the atoms within the thin skin depth from the cloud surface.
The origin of the effect can be traced to original equations of
radiative transfer (see eq.~(\ref{simplified})). For instance, 
in the absence of
velocity effects $\rho_s$ is proportional to the integral over the whole
line of sight (see eq.~(\ref{Y})).

\subsection{Thin Velocity Slices of Spectral Line Data}  

Within the VCA technique introduced in LP00 the change of
the spectral index of emission fluctuations with the thickness
of the channel maps was employed to separate the contributions
from density and velocity. In particular, 
to see the {\it thin} slice scaling that reflects the statistics of  turbulent velocities,
the effective window of the velocity
slice
should be narrower than the turbulent velocity dispersion at the 
scale of the study.

The 
thermal broadening (i.e. $\beta$ factor in the previous section) of the line
can be considered as a convolution with the window $W_{slice}$ describing the
channel slicing adopted in the data analysis.
The LP00 study showed that adopting the slice thickness less
than the thermal width $\delta V < \beta$ does not provide new information
about turbulence\footnote{It does provide the information about
the gas temperature, however. This point will be elaborated 
elsewhere.}. In our discussion we shall treat thermal effects as
adding (in quadrature) the minimum width to the data slice,
symbolically ${\delta V}^2+\beta$

Absorption results in appearance of the second
window function $W_{absorption}$. Two cases can take place.

If $v_{ab}^2 < {\delta V^2}+\beta$ the regime is not different from
that discussed in the earlier subsection. Indeed, it is the
absorption that limits the integration and the results
of integration over the slice thickness and over the whole
emission line should be similar. According to previous
discussion, absorption smoothing reveals the underlying
velocity statistics only for $m < 2/3$.
Since 
the thermal broadening is given by the properties
of the media under study and therefore is
fixed, we get the following picture.
If absorption is strong
$v_{ab}^2 < \beta$ making slices thinner will
not provide any additional information about turbulence. 
This is the case when the velocity information
recovered using the VCA is limited if any.

In case  $v_{ab}^2> \beta$ one can get information about turbulence by
making the slice thinner, i.e. decreasing $\delta V$ to
$ {\delta V}^2 < v_{ab}^2-\beta$.
Choosing the thickness of the velocity slice to ensure that
$v_{ab}^2 > {\delta V}^2 + \beta$ we formally return to the regime
very similar to that considered in LP00. Indeed, if the window
function defined by the absorption is unimportant we return to
the case mathematically similar to that in LP00. Naturally
our results obtained in this regime are identical to those in LP00.
Fig.~4 shows what is happening
when we choose thin slicing for self absorbing data. It is easy to
see that in doing so we get back to the regime of thin slices discussed in
LP00. 

As the result we can extend the VCA for self-absorbing clouds
and provide observers with a following recipe:\\
{\it 1. If the integrated line emissivity shows a universal spectrum
with $n=-3$, this testify that the density statistics is steep. In this
case thin slice regime would provide one with a spectrum, which power-law
index is $-3+m/2$.\\
2. If the integrated line emissivity shows a spectrum which is shallower
than $-3$, this index should be identified with the power-law index $n$ of
the underlying 3D density field. The thin slice regime in this case provides
the index $n+m/2$.}

It is clear that in the case of shallow density the VCA recovers both 
3D velocity and density spectra, while in the case of steep density the
VCA would normally recover just velocity spectral index. Potentially,
one may attempt to remove the universal statistics in this case, but
this may be very challenging in practical terms. It is also evident that
for subsonic turbulence thin slice regime is not attainable.

\section{Accomplishments and Qualitative Understanding of Results}

Let us start with an explanation what has been done so far
in different parts of the present paper. We started
with introducing radiative transfer in \S 3 and obtaining
general expressions for the statistics of intensity in
the presence of the radiative transfer. We showed that
absorption plays the role of an additional window function
in those expressions. Then in \S 4 we discuss the two 
limiting and observationally important cases. One is
related to the way how most present day observations 
of fluctuations are performed, namely, through measuring
fluctuations of the integrated intensity. In this
situation we show that the information about the underlying
statistics in most cases is lost. We show, however, that
the information can be recovered if fluctuations in velocity
slices are analyzed. These are two major results of the paper.

A good deal of the paper and most of the Appendixes deal with
the statistics of the fluctuations without absorption. This
was necessary because the earlier machinery was formulated in
terms of power spectra, while the transformation from the
PPV power spectra to correlation functions happen to be
not trivial. In terms of our earlier work, this not only
provides an independent check of our results in LP00, but
provides a very important extension of the technique.
For instance, we obtain the predictions for the statistics
of fluctuations along the velocity direction. These fluctuations 
can be studied using instruments with good spectral resolution 
even if the spatial resolution of the instruments is poor.    

With the mathematical machinery in the paper being sufficiently
involved, the issue of qualitative understanding of our results
is of a particular interest. Especially this is important
for our results that may look somewhat counterintuitive.

Consider first the formulae for thin slices. The spectrum
of intensity in a thin slice gets shallower as the underlying
velocity get steeper. To understand this effect consider turbulence
in incompressible optically thin media. The intensity of emission
in a slice of data is proportional to the number of atoms per
the velocity interval given by the thickness of the data slice.
Thin slice means that the velocity dispersion at the scale of
study is larger than the thickness of a slice. The increase
of the velocity dispersion at a particular scales means that
less and less energy is being emitted within the velocity
interval that defines the slice. As the result the image of
the eddy gets fainter. In other words, the larger is the
dispersion at the scale of the study the less intensity
is registered at this scale within the thin slice of spectral
data. This means that steep velocity spectra that correspond
to the flow with more energy at large scales should produce
intensity distribution within thin slice for which the
more brightness will be at small scales. This is exactly
what our formulae predict for thin slices (see also LP00).

The result above gets obvious when one recalls that the largest
intensities within thin slices are expected from the regions that
are the least perturbed by velocities. If density variations are
also present they modify the result. When the amplitude of density
perturbation becomes larger than the mean density, both the
statistics of density and velocity are imprinted in thin slices.
For small scale asymptotics that we are mostly interested in 
the paper, this happens, however, only when the density spectrum
is shallow, i.e. dominated by fluctuations at small scales.

Understanding of the results for integrated spectral line in the presence
of absorption is a bit more involved. Again, it is advisable to consider
 incompressible flows first. If no absorption present the integrated
spectral line images are not affected by velocity fluctuations. Indeed,
the velocity field just redistributes atoms along the line of
sight and this redistribution cannot affect the total intensity in
the absence of absorption. 

If absorption is present, the fact that velocity redistributes atoms
along the line of sight causes fluctuations of the integrated intensity.
Two effects, however, are present simultaneously as the velocity
field at a particular scale spreads atoms in PPV space. It is easy
to see that the most absorption is expected is when the atoms have
the same velocities. Therefore velocity dispersion provided by
turbulence decrease the absorption and therefore increases the signal.
However, the decrease of the optical thickness makes the media
to behave more like
its optically thin counterpart, in which the velocity fluctuations
are averaged out. In other words, while the mean level of intensity increases
as the result of turbulence, the contrast of the fluctuations caused
by the velocity fluctuations decreases. The range of the universal
asymptotics $K^{-3}$ is the range at which the two effects {\it
exactly} compensate each other.

Density fluctuations get important only when their amplitude gets
larger than the mean density. This is quite analogous to the case
of thin slices discussed above. The outcome is also analogous, namely,
for the small scale asymptotics only density with shallow spectrum 
is important. Density with steep spectrum may affect the large scale
asymptotics.

\section{Discussion}

\subsection{ Description of turbulence in absorbing media}

In LP00 we provided the statistical description of spectral line
data cubes obtained via observations of turbulent media. In the 
present paper we generalized this description for the presence of
the absorption effects.

If we compare our present treatment of PPV statistics with that in LP00 
we see that it differs 
in terms of the statistical tools used. As the absorption takes place in
real and not Fourier space, the use of structure functions
is advantageous. In Appendix~D we re-derive LP00 results for
optically thin media using structure functions.
Our present treatment
extends the VCA technique for studies of turbulent absorbing 
media. How good is our treatment of absorption?

Above we used 1D model of radiative transfer. We would, 
however, argue that it is adequate
for our purposes. Indeed, our aim is to study 
velocity and density fluctuations
before the effects of optical depth distort the underlying power-law 
spectra. On the contrary, when the 3D radiative transfer gets important
we expect the non-linear 
flattening of the observed emissivity spectra.

For the sake of simplicity in our treatment we
made major calculations assuming that the fluctuations within PPV data
cubes are Gaussian. However, we showed that the structure
of our expressions does not depend on this assumption. Whatever
is the statistics of emissivity fluctuations in PPV space the
absorption introduces a new window function that truncates the
velocity range over which the integration is provided. Possible
changes in the window function arising from non-Gaussian fluctuations
are expected to result in the unimportant changes of the numerical
prefactors of the expressions.

We can test our result in the limiting cases of high and low absorption.
In the latter case our formulae can be expanded and we get 
the LP00 result (see also Appendix~C) for $\alpha \rightarrow 0$.
For high absorption, it is only very small fluctuation that
are still in the linear regime. Those small fluctuations are likely
to be Gaussian.

There are a number of issues that are relevant to studies
 of turbulence in both absorbing and not absorbing media.
One of them is importance of velocity-density correlations.
While the issue, in general, require further research, the
case of velocity-density correlations has been studied
analytically in LP00 and numerically 
(Lazarian et al. 2001, Esquivel et al. 2003). Our discussion
above reveals that when the effective thickness of the slice
is less than window function given by absorption $W_{absorption}$
the effects of absorption are unimportant. For such slices,
if they are thin, the effect of velocity-density correlations
was shown to be marginal.

For the sake of simplicity both in our present paper 
and in LP00 we assumed that the
emissivity is proportional to the first power of density. 
If the dependence is non-linear, e.g. proportional to 
$n^2$ as in the case of H$_\alpha$ emission,
for small amplitude perturbations it is possible to show in the
linear terms dominate and therefore our results
above are applicable (see discussion in Cho \& Lazarian 2002b)\footnote{
One may wonder whether complications arise at
large scales. We believe that this is unlikely for steep densities.
For instance, for H$_{\alpha}$ the emissivity
is proportional to density squared. The correlation functions of
density $2\langle (\delta \rho_1)^2 (\delta \rho_2)^2 \rangle$ correspond
to a steeper spectra and therefore their contribution within thin slices
is sub-dominant.}.

An additional simplification shared by this work and LP00 is
that the thermal velocity of 
media is assumed constant. As discussed above thermal velocities affect the
effective window function over which the turbulent velocities are smeared.
As the result if observations sample regions with different temperatures
of gas, the thermal smearing will provide higher weighting 
to the regions where thermal velocity
is less than the slice thickness. This stays true in the
presence of absorption. 

Another non-linearity disregarded in the treatment above 
as well as in LP00 is related to
the non-linearities in the Galactic rotation curve. The effects of
the rotation curve shear were studied in Esquivel et al (2003).
There it was shown that effects of the shear can be ignored unless the
galactic shear and the shear provided by the eddies are comparable.
The latter is a very artificial situation as the large scale shear
in most cases produces turbulence. For Kolmogorov turbulence the local shear
$v_l/l$ scales as $l^{-2/3}$. Magneto-rotational instability (see
Chandrasekhar 1967, Balbus \& Hawley 1995) ensures that the global 
galactic shear
creates turbulence. If the shear provides marginal influence, the higher
order corrections provided by the non-linearity of the shear are 
also negligible.

Our treatment above assumes that the absorption happens within
the emitting gas. What will happen when the radiation
source is outside the turbulent cloud?  
If turbulence is studied using {\it absorption} lines the registered
intensity is 
\be
I=I_0\exp\left[-\int_o^L \alpha(z)n(z)\phi(z)dz\right]
\ee
where $I_0$ is the intensity coming from the external source.
The easiest way is to correlate logarithms of
$I/I_0$. For constant $\alpha$ 
these quantities are evidently proportional to the 
integrals of emissivities and the analysis in LP00 is directly
applicable. 

Is it always true that we have to deal with velocity fluctuations?
If the velocity turbulence gets damped, while density statistical
structure persists, e.g. in the form of entropy fluctuations,
 the effects of velocity modification 
get marginal\footnote{An important case of this regime is the viscosity
damped MHD turbulence reported in Cho, Lazarian \& Vishniac (2002c). 
This new type of MHD turbulence produces small scale magnetic fluctuations
through stretching of magnetic field by marginally damped eddies (
Lazarian, Vishniac \& Cho 2003). The slowly evolving
fluctuations of magnetic field
result in the fluctuations of density that are essentially stationary
(Cho \& Lazarian 2003ab).}.

\subsection{Interpretation of Observations}

In a number of instances the power-law spectra of interstellar
turbulence were reported for the spectral line data. They 
include HI (Green 1993, Stanimirovic et al. 1999) 
CO in molecular clouds (Stutzki 1998), H$_{\alpha}$ in 
Reynolds layer (Minter \& Spangler 1996),
HI absorption (Deshpande, Dwarakanath \& Goss 2000).
In all the cases the reported index was close to $-2.7$, which is
really amazing as in all those cases very different quantities
were measured\footnote{The spectral index obtained in Stanimirovic et al.
(1999) was somewhat steeper, namely, $-3$. However, it was shown
in Lazarian \& Stanimirovic (2001) that this difference stemmed from
the difference in the slice thickness. As slice thickness was reduced
the index became $\approx -2.8$ over the whole range of scales under
study.}. In comparison, fluctuations of synchrotron emission
and starlight polarization show very different indexes, although
underlying turbulence is likely to be Kolmogorov (Cho \& Lazarian 2002).

Results of the earlier applications of VCA to the HI data
can be summarized as follows:
I). the observed scaling of 2D power spectra within thin slices is
consistent with arising from velocity caustics; II).
the density power spectrum is steep at least in the case of
Small Magellanic Cloud data; III). HI data is consistent
with the velocity spectral index being approximately Kolmogorov.

The interpretation of H$_\alpha$ data may be done along the same line of
reasoning. Earlier we argued that non-linearities
in the H$_\alpha$ emissivities are not important for
studies of small-scale fluctuations. This would also indicate the
correspondence with Kolmogorov spectrum of turbulence of velocity.

Consider now HI absorption. Instead of the over-density 
the absorption measures deal with the ratio of density over
temperature $n(x)/T(x)$. Both quantities fluctuate and
{\it a priori} one expects the spectrum to differ from that of HI emission. 
However, 
Deshpande et al. (2000) found that the index of the absorption spectrum in
slices of their data is similar to that Green (1993), but over the 
range from 3~pc to 0.07~pc. However, at scales of approximately
0.2~pc a change of the index of the power spectrum was observed.
Therefore the observed power law covers approximately one order of
magnitude scales from 3~pc to 0.2~pc. If at those scales turbulent
velocities are larger than thermal velocities, the slices used
may be thin and the coincidence of the spectral index in the emission
and absorption would mean the continuation of the turbulent cascade to
small scales. This, however, may be a bit surprising as the velocity
turbulence at those scales is likely to be damped through viscosity
by neutrals unless the turbulence is generated at very small scales.

To determine what is going on Deshpande et al. (2000) used VCA prescriptions 
to evaluate the effects of the velocity and did not detect the change
of the spectral index as he changed the thickness of the data slice.
 Deshpande et al. (2000) could not see the changes of the spectral
index and this can be interpreted that the fluctuations arise exclusively
from density variations with the shallow spectral index ($\sim -2.7$).
 
If further research confirms that the density spectrum
is indeed shallow this may be the signature of the magnetic energy 
cascade simulated in  Cho, Lazarian \& Vishniac (2002), Cho \&
Lazarian (2003) and described in Lazarian, Vishniac \& Cho (2003). 
In this new regime of turbulence fluctuations of density arise from
fluctuations of magnetic field, while velocity fluctuations are 
marginal. The expected power spectrum of density scales $k^{-3}$ and is
roughly consistent with the observations. Note, that such
a shallow spectrum of density according to Deshpande (1999) can 
explain the detected mysterious tiny scale structures 
in HI (see Heiles 1998).It is obvious that a further study of the
statistics of HI fluctuations at small scales is necessary.
Synergy of different techniques can be called for. For instance,  
modified velocity centroids (Lazarian \& Esquivel 2003) can provide
an independent test whether velocity fluctuations are important.

All the data above was obtained by studies of fluctuations within
velocity slices of spectroscopic data. On the contrary,
CO data by Stutzki et al.(1998) present the power-spectrum of fluctuations
obtained via integrating the intensity over the whole emission line. 
According to Falgarone et al. (1998) the turbulent velocities even
at the smallest scales are still supersonic. Thus we expect to see
a velocity modification of the spectrum.
Stutzki et al. (1998)  get the index of order $-2.8$. Accepting
possible criticism that in the case of HI we assumed that the
deviation of $-2.7$ from $-3$ is meaningful and informative,
we may speculate that the measured CO spectral index 
are close
to the universal asymptotics of $-3$ predicted in
the paper above. We may speculate that the flattening of
the spectrum compared to the expected $-3$ may be caused in
the case of CO by instrumental noise (this possibility was
mentioned to us by our referee). 
 
Alternatively, one expects to see the power index $n+m/2$ if the
density spectral index $n>-3$. If we want to account for $-2.8$
index we would have to accept not only that the density is
shallow, but also that the velocity is more shallow
that the Kolmogorov value. More studies utilising our present theoretical
insight are clearly necessary.

What are the goals that can be achieved by studying properties
of interstellar turbulence?
Establishing turbulence spectra by applying VCA to different emission
and absorption lines can help
to answer fundamental questions of the interstellar physics. 
For instance, it can answer the question whether the turbulence
in molecular clouds is a part of a bigger cascade from the large
scales or it is locally generated. It can also answer the question
to what extend the self gravity is important in determining the
spectra of fluctuations at different scales.

In addition, VCA states that the intensity of fluctuations
measured in a velocity channel depends on whether the channel
width is larger or smaller than the thermal velocity dispersion
of the atoms. This potentially allows to find the distribution of
temperatures of emitters along the line of sight. This may provide
an independent testing of theories of the ISM thermal structure
and answer the question of the role of thermal instability for
the ISM (see Heiles \& Troland 2003).   
 
\subsection{Range of Applicability and Numerical Testing}

The VCA technique is the most efficient for restoring the underlying
spectral indexes when the turbulence obeys the power-law
and the turbulent velocities are larger than the thermal velocities
of atoms. While the first requirement is usually fulfilled in interstellar
turbulence (see Stanimirovic \& Lazarian 2001), numerical simulations
provide a limited inertial range if any. This presents problems when
the testing is involved. For instance, in Lazarian et al. (2001)
and Esquivel et al. (2003) the spectral slope was corrected to fit
a power-law. While the power-law requirement is not an essential
constraint for most of the interstellar situations, the ratio of the 
turbulent to thermal velocity may be a constraint. To use the
VCA for the subsonic turbulence one should use the heavier species,
which thermal velocities are less than that of turbulence if the
thin slice asymptotics is intended to be observed. Note, that 
in case of turbulence  that is supersonic at large scales 
this potentially limits the range of the small scales which can
be studied through the VCA technique. Indeed, at some small scale
the turbulence gets subsonic and the slope of the emissivity
fluctuations within a slice should change\footnote{This change can be used
to determine at what scale turbulent and thermal velocity are equal.
If the dispersion of turbulent velocities is known at the large
scales, this allows to measure the temperature of gas. If, alternatively,
the temperature of gas is known, this allows to calibrate the power
spectra in terms of absolute values of turbulent velocities.}.

Another problem of numerical testing is the appearance of shot noise
related to the limited number of data points available through simulations.
Unless understood, this noise produces a bias in the power spectral indexes 
obtained through synthetic observations based on the numerically generated
data cubes. This problem was analyzed in Esquivel et al. (2003), where it
was concluded that it is not important for actual observations where the
number of emitters is essentially infinite. When absorption is taken
into account our preliminary analysis shows that the testing is becoming
even more challenging and requires piling up many numerical data cubes in
order to obtain adequate statistics.

The non-trivial issues above may cause confusion. For instance, in 
 Miville-Deschenes et al. (2003a) the utility of the VCA was
questioned. We have to reiterate that the effects that the authors
encountered, e.g. shot noise, are related to the synthetic data they 
used and not
to the actual observational data. None of the problems they discuss
was present when supersonic turbulence of SMC was studied in
Stanimirovic \& Lazarian (2001). 

In another paper Miville-Deschenes et al. (2003b) mention limitations
of the VCA related to the thermal motions of atoms. It is pointed
out there that the thermal line width of atoms can entail asymptotic
behavior that can be mistakenly identified with the actual saturation
of the power-law index in a thin slice. In fact, the observed
saturation in the two cases is different. If the thermal line width gets 
larger than the thickness
of the slice, further decrease in the slice thickness does not change
the shape of the slice emissivity spectrum. It consists of two parts,
a shallow part corresponding to the scales at which turbulent velocity is 
larger
than $v_{thermal}$ (thin slice) and steeper part corresponding to the 
scales at which
the turbulent velocity is larger than $v_{thermal}$ (thick slice). On the 
contrary, if the turbulence is supersonic on the scales under the study, the
change of the slice thickness makes the change from thin to thick slice to 
move
to smaller and smaller scales. In other words, the slice is thin at large 
scales
and thick at small scales with the border between these two regime moving
as the thickness of the slice changes, provided that the experimentally 
determined
thickness of the slice is larger than the thermal broadening of the 
line\footnote{The good news is that by varying the thickness of the 
velocity slices
on can find the ratio of the thermal and turbulent velocities (LP00).}.

Such a study has not been done in Miville-Deschenes et al. (2003b). 
Neither they
give estimates of the thermal and turbulent velocities
of gas in the region Ursa Major Galactic cirrus they studied. Instead 
they pointed
out to the discrepancy between the spectral index obtained using the 
velocity centroid
analysis and those the VCA would give if the asymptotic regime 
{\it were not} due
to thermal broadening. By itself, this shows important synergy of 
different techniques.
Unfortunately, the authors did not check whether their centroids are 
dominated by
velocity or density. The application of the criterion obtained in LE03 to the 
data seems essential for answering this question. High degree of 
correlation between
the averaged density and the averaged centroids maps as well as a close 
similarity
of the spectra deduced from averaged density and the centroid maps  
calls for further
testings of what the velocity centroids measure velocity in the particular case 
(see LE03). A study in Esquivel \& Lazarian (2004) 
testifies that velocity centroids
are dominated by density if magnetic turbulence has Mach numbers higher
that 2. This is also suggestive that centroids obtained for the
hypersonic turbulence in HI are
dominated by density. 

We would like to stress that the importance of 
quantitative understanding of the statistics of spectral line data cubes 
goes far beyond the particular recipes of the VCA technique. LP00 and
this paper provides for the first time the machinery suitable for 
dealing with the statistics of spectral line data. This machinery should
provide advances not only when analysis of spectra in channels is involved,
but also suggests new techniques. For instance, our treatment of velocity
correlations in $v$ direction allows a new way of measuring the spectral
index of velocity fluctuations.

\section{Summary}

In this paper we have studied effects of absorption on the statistics
of spectral line data. In particular, we discuss
how the 3D velocity and density statistics of turbulence
can be recovered from the spectral line data. We subjected
to scrutiny two regimes of studying turbulence, namely, (a) when
integrated spectral line intensity is used and  (b) when the
channel maps are used. For our treatment we assumed the power-law
underlying velocity and density statistics. Our results are summarized
below.

I. For integrated spectral line data we obtained:

1. Absorption introduces non-linear effects that distort the power
law behavior of measured intensity. The non-linear 
effects are marginal when fluctuations are studied at sufficiently 
small scales.

2. While in the absence of absorption
the velocity effects were washed out for the integrated
spectral data, absorption makes velocity modification prominent.
The recovery of the underlying velocity spectrum from integrated
spectral data is non-trivial and sometimes impossible. For a range
of scales the resulting power spectrum of intensity is
caused by velocity fluctuations but
does not depend on the spectral index of velocity fluctuations.

3. Density spectra can be recovered in the limit of small scale
asymptotics that we mostly deal here only when the density
has a shallow spectrum, i.e. is dominated by fluctuations at
small scales, or at scales where turbulence gets subsonic. 

II. For spectral data in channel maps we obtained:

1. If slices are sufficiently thin we recover thin slice asymptotics
obtained earlier in LP00. However, the choice of slice thickness
is limited by absorption, which introduces the minimal thickness of
the slice for which non-linearity is negligible.

III. We incorporated our results into the VCA technique intended
for the observational studies of turbulence. The detailed description
of the spectral line data cube statistics that we obtained should
enrich the set of tools available for advancing quantitative insight into
interstellar turbulence.

\acknowledgments

Part of this work was initiated while one of us (AL) enjoyed the
stimulating environment of the 
 Astrophysical Turbulence Program 2000 at the Institute
for Theoretical Physics at the University of California at Santa
Barbara. AL also acknowledges the Visiting Professor position
at Ecole Normale that enabled some of the discussions helpful for
this research.
Research of AL is supported by the NSF Grant  AST-0307869 and by
NSF Center for Magnetic Self-Organization in Laboratory and Astrophysical 
Plasmas, while research of DP is supported
by the Natural Sciences and Engineering Research Council of Canada.

\appendix

\section{Statistics of velocity and density in XYZ coordinates}

{\it Statistics of a finite-size cloud.}
If we deal with a cloud of size $S$ the turbulence has a maximal
scale.
In the absence of additional physics the size $S$ also serves as
structure function cutoff scale $D(r) \sim D(\infty) r^m/(r^m+S^m)$.
We notice that if this law is 
adopted for $D_{LL}(r)$, z-component structure function in the case of
solenoidal turbulence is
\begin{eqnarray}
D_z({\bf r}) &=& D(\infty) \frac{r^m}{r^m+S^m} \left( 1+  
\frac{m/2}{1+(r/S)^m}(1-\cos^2\theta)\right), ~~~~~ 
\cos\theta={\bf r} \cdot {\bf \hat z }/r \\
D_z(\infty)&=&D(\infty)=C  S^m
\label{struc_intro}
\end{eqnarray}
We shall further assume that the turbulent velocity field is Gaussian,
in which case its properties are fully determined by the two point 
probability distribution function $\mathrm{P}(u_1,u_2)$,
where $u_1=u_z({\bf x}_1), \; u_2=u_z({\bf x}_2)$.  
Using variables $u=u_1-u_2$, $u_+=(u_1+u_2)/2$ this function is conveniently
expressed (we write
down the distribution of z-component of 
the velocity field $u({\bf x})$ only) as
\begin{eqnarray}
{\mathrm P} (u_1,u_2) &=& P(u,u_+) \label{Pu1u2} \\ \nonumber 
&=&\frac{1}{\pi \sqrt{2 D_z(\infty) - D_z({\bf r}) } \sqrt{D_z({\bf r})}}
\exp\left[-\frac{u^2}{ 2 D_z({\bf r})}\right]
\exp\left[-\frac{u_+^2}{ D_z(\infty) - D_z({\bf r})/2}\right]
\end{eqnarray}
Validity of eq.~(\ref{Pu1u2}) is immediately checked by observation that
$u,u_+$ are uncorrelated $\langle u u_+ \rangle \sim \langle u_1^2 \rangle
- \langle u_2^2 \rangle = 0$ Gaussian quantities with dispersions
\begin{eqnarray}
\langle u^2 \rangle &=& \langle (u_1-u_2)^2 \rangle = D_z({\bf r}) \nonumber \\
\langle u_+^2 \rangle &=& \frac{1}{4} \langle (u_1+u_2)^2 \rangle = 
\frac{1}{4} \left[2 \langle u_1^2 \rangle + 2 \langle u_2^2 \rangle
- \langle (u_1-u_2)^2 \rangle\right]
= \frac{1}{2}\left[D_z(\infty)-D_z({\bf r})/2\right] ,
\nonumber
\end{eqnarray}
and the Jacobian of transformation between $u,u_+$ and $u_1,u_2$ is unity.

{\it Shallow density spectrum}\\$u_1=u_z({\bf x}_1), \; u_2=u_z({\bf x}_2)$ 
When statistical properties of the density fluctuations  
dominated by short wavelengths, we use  power-law correlation functions
of {\it over-density}:
\begin{equation}
\xi(r)= \langle \rho \rangle^2 
\left(1 + \left( {r_0 \over r} \right)^\gamma\right), ~~~~~ \gamma > 0~~~.
\label{Appeq:xi}
\end{equation}
Note, that the power-law part of the correlation corresponds 
to the 3D power-law spectrum 
$\propto k^n$, where  $n=\gamma-3$ which for $\gamma>0$ means that 
$n$ is shallow, i.e. less than $-3$.

{\it Steep density spectrum}\\
To describe the density statistics 
 for steep (i.e. $n<-3$) power spectrum one should
use structure function description. For $\gamma<0$ 
we write
\be
\xi(r)=\frac{1}{2} d(\infty)[1-d(r)/d(\infty)]~~~,
\ee
where $d(r)$ is a structure function  of density
given by Eq.~(\ref{2})

Real world structure functions do not grow infinitely and therefore
we have to introduce a cutoff at some large scale. If the 
cut-off happens at $r_c$ then for the long-wave dominated turbulence 
($\gamma<0$)
\be
d(r)=d(\infty){r^{-\gamma} \over r^{-\gamma}+r_c^{-\gamma}}~~~,
\label{structure}
\ee
and the correlation function is
\be
\xi(r)=\frac{d(\infty)}{2}{r_c^{-\gamma} \over r^{-\gamma}+r_c^{-\gamma}}~~~.
\label{cor-structure}
\ee
For sufficiently small $r\ll r_c$ eq.~(\ref{cor-structure}) gives
\be
\xi(r)\approx \frac{1}{2} d(\infty) (1-[r/r_c]^{-\gamma})
\label{xinew}
\ee
Note that in eq.~(\ref{xinew}) 
$r_c [\langle \rho \rangle^2/d(\infty)]^{-\frac{1}{\gamma}}$
plays the role of $r_0$ in eq.~(\ref{Appeq:xi}).

When using both correlation and structure functions
one essential difference should be kept in mind: while the value
of the structure function at some scale $r_*$, $d(r_*)$,
is determined by the power of
fluctuations at smaller scales $r \le r_*$, the value of the correlation
function $\xi(r_*)$ reflects the integral power at scales $r \ge r_*$.
For the power-law statistics, the correspondent integral contributions
are dominated by the power at scale $r_*$ itself, localizing the information
provided by $\xi(r_*)$ and $d(r_*)$ values.

\section{Statistics of density in PPV}
In this Appendix we calculate some statistical properties of the
velocity space density
\begin{equation}
\rho_s({\bf X},v) = \int_0^S {\mathrm d} z\; \rho({\bf x}) \phi_v({\bf x}),
\end{equation}
where
$\phi_v({\bf x})$ is given by eq.~(\ref{phi})

and random fields $ \rho({\bf x})$ and $u({\bf x})$ over distribution of which
we average are assumed to be uncorrelated.  We assume turbulent velocity field
to be described by the Gaussian two-point probability distribution function
given in eq.~(\ref{Pu1u2}).
We shall also consider statistical properties of the density
distribution in the galactic coordinates to be homogeneous:

The mean PPV density is given by
\begin{eqnarray}
\langle \rho_s({\bf X}_1, v_1)\rangle &=&
\int_0^S dz_1 \; \langle \rho({\bf x}_1)\rangle \;
\langle \phi_{v1}({\bf x}_1) \rangle = \nonumber \\
& = & \frac{\bar\rho}{\left( \pi \left[D_z(\infty)+2\beta\right]\right)^{1/2}}
\int_0^S dz_1 \; 
\exp\left[-\frac{(v_1-v_{gal,1})^2}{D_z(\infty)+2\beta}\right]
\label{AppB:mean}
\end{eqnarray}
and the two point correlation function is
\begin{eqnarray}
\langle \rho_s({\bf X}_1, v_1)
\rho_s({\bf X}_2,v_2)\rangle &=& \int_0^S dz_1 \int_{0}^{S} dz_2 \;
\langle \rho({\bf x}_1)\rho({\bf x}_2)\rangle \;
\langle \phi_{v1}({\bf x}_1) \phi_{v2}({\bf x}_2) \rangle = \nonumber \\ 
&=&\frac{1}{2 \pi}
\int_{-S}^S {\mathrm d}z 
\int_{|z|/2}^{S-|z|/2} {\mathrm d}z_+
\; \frac{\xi({\bf r})}{[D_z({\bf r})+2\beta]^{1/2}}
\exp\left[-\frac{(v-v_{gal})^2}{2 (D_z({\bf r})+2\beta)}\right] \nonumber \\ 
&\times& \frac{\sqrt{2}}{ [\beta+D_z(\infty)-D_z({\bf r})/2]^{1/2}}
\exp\left[-\frac{(v_+-v_{gal,+})^2}{\beta+D_z(\infty)-D_z({\bf r})/2}\right]
\label{AppB:2point}
\end{eqnarray}
Particular case of eq.~\ref{AppB:2point} is the second moment
\begin{eqnarray}
\langle \rho_s({\bf X}_1, v_1)^2\rangle & = & \int_0^S dz_1 \int_{0}^{S} dz_2 \;
\langle \rho({\bf X}_1,z_1) \rho({\bf X}_1,z_2)\rangle \;
\langle \phi_{v1}({\bf X}_1,z_1) \phi_{v1}({\bf X}_1,z_2) \rangle = \nonumber \\ 
&=&\frac{1}{2 \pi}
\int_{-S}^S {\mathrm d}z 
\int_{|z|/2}^{S-|z|/2} {\mathrm d}z_+
\; \frac{\xi(|z|)}{[D_z(|z|)+2\beta]^{1/2}}
\exp\left[-\frac{v_{gal}^2}{2 (D_z(|z|)+2\beta)}\right] \nonumber \\ 
& \times & \frac{\sqrt{2}}{ [\beta+D_z(\infty)-D_z(|z|)/2]^{1/2}}
\exp\left[-\frac{(v_1-v_{gal,+})^2}{\beta+D_z(\infty)-D_z(|z|)/2}\right]
\label{AppB:sigma}
\end{eqnarray}
Here we have used our standard notation $v=v_1-v_2$, $v_+=(v_1+v_2)/2$,
$z=z_1-z_2$, $z_+=(z_1+z_2)/2$, ${\bf r}={\bf x}_1 - {\bf x}_2$,
$v_{gal}=v_{gal}({\bf x}_1)-v_{gal}({\bf x}_2)$ and
$v_{gal,+}=[v_{gal}({\bf x}_1)+v_{gal}({\bf x}_2)]/2$.

We observe that the  density in the velocity space is statistically inhomogeneous
which  is reflected first of all in a residual dependence of the quantities in
eqs.~\ref{AppB:mean}-\ref{AppB:sigma}
on the absolute velocity $v_1$ or $v_+$ and not only on velocity difference $v$. 
In particular, eq~\ref{AppB:mean} describes the mean velocity profile of the
of the density, which is not, in general, uniform.

There can also be dependence of the statistics on sky coordinates ${\bf X}$, if
the regular flow pattern $v_{gal}({\bf x})$ is complex. We shall not consider
this possibility, restricting our attention either to $v_{gal}$ described 
by linear shearing pattern or the case where
regular flow can be neglected altogether as is the case for isolated clouds.

Expressions eqs.~\ref{AppB:mean}-\ref{AppB:sigma} are quite complex.
They are significantly simplified in several astrophysically important cases.
First of all, when one considers small enough scales so that
turbulent velocities are smaller than $D_z(\infty)^{1/2}$,  
all exponential terms which contain $D_z(\infty)$ can be taken equal to unity
and we have
\begin{eqnarray}
\langle \rho_s({\bf X}_1, v_1)\rangle &=& const \\
\langle \rho_s({\bf X}_1, v_1)^2\rangle &=& const\\
\langle \rho_s({\bf X}_1, v_1)\rho_s({\bf X}_2,v_2)\rangle &\propto &
\int_{-\infty}^\infty {\mathrm d}z \left( 1-\frac{|z|}{S}\right)
\; \frac{\xi({\bf r})}{[D_z({\bf r})+2\beta]^{1/2}}
\exp\left[-\frac{(v-v_{gal})^2}{2 (D_z({\bf r})+2\beta)}\right]
\label{AppB:2point_inf}
\end{eqnarray} 
This equations are applicable when the velocity line is extended compared with
the turbulent scales under study which is the case, for example,
 for the 21cm emission of interstellar HI.
Henthforth we shall not follow precise values
of the proportionality factors,
which depend on the extend of the line-of-sight integration $S$ 
and the details of the integral cut-off. For the analysis of the
scaling laws at small separations $R$ the boundary effects can be neglected
and the range of integration extended to infinity.

In another limit, when dealing with lines from individual clouds, one can
consider regular flow to be absent. In this case
\begin{eqnarray}
\langle \rho_s({\bf X}_1, v_1)\rangle & = &
\frac{\bar \rho S}{\sqrt{\pi} \left[D_z(\infty)+2\beta\right]^{1/2}}
\exp\left[-\frac{v_1^2}{D_z(\infty)+2\beta}\right]
\label{AppB:mean_cloud} \\
\langle \rho_s({\bf X}_1, v_1)\rho_s({\bf X}_2,v_2)\rangle &\propto&
\frac{{\bar \rho}^2 S}{ [D_z(\infty)+\beta]^{1/2}}
\exp\left[-\frac{v_+^2}{D_z(\infty)+\beta}\right] 
\label{AppB:2point_cloud} \\
&\times&\int_{-S}^S {\mathrm d}z \left(1-\frac{|z|}{S}\right)
\; \frac{\xi({\bf r})/\bar \rho^2}{[D_z({\bf r})+2\beta]^{1/2}}
\exp\left[-\frac{v^2}{2 (D_z({\bf r})+2\beta)}\right] \nonumber \\
\langle \rho_s({\bf X}_1, v_1)^2\rangle &\propto& 
\frac{{\bar \rho}^2 S^2}{ [D_z(\infty)+\beta]^{1/2}}
\exp\left[-\frac{v_1^2}{D_z(\infty)+\beta}\right]
\label{AppB:sigma_cloud}
\end{eqnarray}
Again, for small separations $R$, the integration can be extended to an
infinite range, $S\to\infty$.

An image of an individual cloud in PPV space has an extension 
$\sim D^{1/2}(S)$. This stems from the fact that for an individual 
cloud the motions on the large scale determine the velocity
dispersion. The assumed distribution, e.g. the Gaussian distribution,
 of large scale cloud velocities
is present for an ensemble of clouds, while for an individual cloud
under study its extension in the velocity space is a fixed realization
from this distribution. Typically, for an individual cloud,
the mean density is fixed. 
As the result the terms like 
$\exp[-v_1^2/2(\beta +D(S))]$
in Eqs.~(\ref{AppB:mean_cloud}-\ref{AppB:sigma_cloud}) represent
numerical factors and will be omitted in our further discussion.

\section{Correlation and structure functions in PPV}

The study of the 3D correlation function in PPV space has been performed
in LP00, with an emphasis on the case of galactic turbulence 
characterized by the presence of the coherent flow due to galactic rotation.
These results follow from eq.~(\ref{AppB:2point_inf})
which can be written in the
form equivalent to eq.~(C1) of LP00 if temperature broadening is taken to be
small (or considered later as a part of the velocity slice width):
\begin{eqnarray}
\tilde \xi_s({\bf R},v) &=& \tilde\xi_v{\bf R},v)+
\tilde\xi_\rho({\bf R},v) \\ \nonumber
\tilde\xi_v({\bf R},v) &\propto& 
\int_{-\infty}^\infty {\mathrm d}z 
\; \frac{\bar\rho^2}{D_z({\bf r})^{1/2}}
\exp\left[-\frac{(v-v_{gal})^2}{2 D_z({\bf r})}\right] \\ \nonumber
\tilde\xi_\rho({\bf R},v) &\propto& 
\int_{-\infty}^\infty {\mathrm d}z 
\; \frac{\tilde\xi({\bf r})}{D_z({\bf r})^{1/2}}
\exp\left[-\frac{(v-v_{gal})^2}{2 D_z({\bf r})}\right]
\end{eqnarray}

In this Appendix we present complimentary set of results for the PPV
structure functions, suitable for description of the gas confined
to isolated clouds. 
Our starting point is the eq.~(\ref{AppB:2point_cloud}).
We shall be interested only in the small
scale $r/S \ll 1 $ behaviour, for which we use simplified expression
\begin{eqnarray}
\tilde\xi_v({\bf R},v) &\propto& 
\frac{{\bar \rho}^2 S}{ D_z(S)^{1/2}}
\int_{-S}^S {\mathrm d}z \;
\frac{1}{D_z({\bf r})^{1/2}}
\exp\left[-\frac{v^2}{2 D_z({\bf r})}\right] \\ 
\tilde\xi_\rho({\bf R},v) &\propto& 
\frac{{\bar \rho}^2 S}{ D_z(S)^{1/2}}
\int_{-S}^S {\mathrm d}z \;
\frac{\tilde\xi({\bf r})/\bar\rho^2}{D_z({\bf r})^{1/2}}
\exp\left[-\frac{v^2}{2 D_z({\bf r})}\right]
\label{AppC:xi_basis}
\end{eqnarray}
where boundary inhomogeneous effects are neglected.
The quantity $\bar \rho_s \approx {\bar \rho} S/D_z(S)^{1/2}$
is the approximate 
mean density in PPV space. At small amplitudes of turbulence velocity one
should restore the thermal term in this expression, 
$\bar \rho_s \approx {\bar \rho} S/(D_z(S)+2\beta)^{1/2}$

Observations of interstellar turbulence testify that the 
emissivity obeys power-laws (see Armstrong, Rickett \& Spangler 1995,
Lazarian 1999, CLV02). This makes it natural to assume
that the underlying velocity and density statistics is also
power-law. Power laws for the statistics of turbulence
are also expected from theoretical considerations (see CLV02). 
In what follows we shall use normalized
variables, which for a cloud of extend $S$ would amount to using
spatial coordinates normalized by $S$ and the velocities normalized
by the velocity dispersion at the cloud scale $CS^m$.
For simplicity, we shall also write $D_z(r)=C r^m$, ignoring non-critical
additional angular dependence exhibited by pure solenoidal or potential
flows.
Asymptotics for $\tilde d_v(R,0),~\tilde d_\rho(R,0),~R/S \to 0$, as well as
$d_v(0,v),~d_\rho(0,v),~v^2/CS^m \to 0$, follow
from analysis of the integral~(\ref{AppC:xi_basis}). Namely
\begin{eqnarray}
\tilde d_v(R,0)&=&2\left[\xi_v(0,0)-\xi_v(R,0)\right] \propto \nonumber \\
&\propto& \frac{{\bar \rho}^2 S^2}{ D_z(S)}
\int_{-1}^1 {\mathrm d}z \;
\left[\frac{1}{z^{m/2}} -
\frac{1}{[(R/S)^2+z^2]^{m/4}} \right]
\propto  \nonumber \\
&\propto& -\frac{{\bar \rho}^2 S^2}{ D_z(S)}
\frac{\sqrt{\pi} \Gamma(m/4-1/2)}{\Gamma(m/4)}
(R/S)^{1-m/2} 
\label{AppC:eq_dvR}\\
\tilde d_\rho(R,0)&=&2\left[\xi_\rho(0,0)-\xi_\rho(R,0)\right] \propto 
\nonumber \\
&\propto& \frac{{\bar \rho}^2 S^2 (r_0/S)^\gamma}{ D_z(S)}
\int_{-1}^1 {\mathrm d}z \;
\left(\frac{1}{z^{\gamma+m/2}} -
\frac{1}{[(R/S)^2+z^2]^{\gamma/2+m/4}} \right)
\propto \nonumber \\
&\propto& -\frac{{\bar \rho}^2 S^2 (r_0/S)^\gamma}{ D_z(S)}
\frac{\sqrt{\pi} \Gamma(\gamma/2+m/4-1/2)}{\Gamma(\gamma/2+m/4)}
\left(\frac{R}{S}\right)^{1-\gamma-m/2}  
\label{AppC:eq_drhoR}
\end{eqnarray}
The latter estimation is valid for $\gamma+m/2 < 1$, i.e for sufficiently
steep density spectra in real space. Kolmogorov density spectrum $\gamma=-2/3$
belongs to this class.  For shallow spectra which have positive $\gamma$,
one should use the correlation function $\xi_\rho$ in place of
$d_\rho$, but one finds (see LP00) that $\xi_\rho(R,0)$ still obeys
the scaling~(\ref{AppC:eq_drhoR}).

Somewhat more complex is the velocity dependence of the structure functions.
For simplicity, we shall describe all the steps in detail for the velocity term.
\begin{eqnarray}
\tilde d_v(0,v)&=&2\left[\xi_v(0,0)-\xi_v(v)\right] \propto \nonumber \\
&\propto& \frac{{\bar \rho}^2 S^2}{ D_z(S)}
\int_{-1}^1 {\mathrm d}z \;
\frac{1}{z^{m/2}} \left[ 1 -
\exp\left(-\frac{v^2}{2 z^m}\right) \right] \nonumber \\
&\propto& \frac{{\bar \rho}^2 S^2}{ D_z(S)} \frac{1}{m p}
\left[ 1 - p \; 2^{-p} v^{2p} \Gamma\left(-p,v^2/2\right) \right]
\label{AppCeq:d_vV}
\end{eqnarray}
where $p=1/m-1/2 > 0$ and $\Gamma$ is incomplete gamma-function.
For small argument, the series expansion of the incomplete gamma-function
is
\begin{equation}
\Gamma[-p,v^2/2] \sim \Gamma[-p] + v^{-2p} \left(\frac{2^p}{p}
+\frac{2^{p-1}}{1-p} v^2 + O(v^4)\right)
\label{AppCeq:Gamma}
\end{equation}
Substituting~(\ref{AppCeq:Gamma})into~(\ref{AppCeq:d_vV}) we see that
the most divergent term in the $\Gamma[-p,v^2/2]$ cancels out, while 
all the residual terms are convergent. The leading behaviour for
the velocity structure function is $d_v(0,v) \sim \Gamma(-p)
v^{2p}$ for $p < 1$ which changes to $d_v(0,v) \sim  \frac{2p+1}{4(p-1)} v^2$
for $p > 1$.  Thus the structure function scaling depends on $m$ for
$m > 2/3$, $d_v(0,v) \sim v^{(2-m)/m}$ and is fixed to $
d_v(0,v) \sim v^2 $ at $m < 2/3$.
\begin{eqnarray}
\tilde d_v(0,v) &\propto& -\frac{{\bar \rho}^2 S^2}{ D_z(S)} \;
\Gamma\left(\frac{1}{2}-\frac{1}{m}\right)
\left(\frac{v^2}{D_z(S)}\right)^{\frac{1}{m}-\frac{1}{2}}, ~~~~~ m > 2/3
\label{AppCeq:d_vVfg23} \\
\tilde d_v(0,v) &\propto& \frac{{\bar \rho}^2 S^2}{ D_z(S)} \;
\frac{1}{2-3m} \; \frac{v^2}{D_z(S)} ~~~~~~~~~~~~~~~~~~,~~~~~ m < 2/3 
\label{AppCeq:d_vVfl23}
\end{eqnarray}
Notably, Kolmogorov value $m=2/3$ is the boundary case here.
One should note that representation~(\ref{AppCeq:Gamma}) is incorrect
for integer $p$, i.e for $m=2/3,2/5,2/7 \ldots $. Series expansion of the
incomplete gamma function is irregular in these cases, which manifests itself
in the appearance of logarithmic terms. For example
\begin{eqnarray}
\Gamma[-1,v^2/2] &\sim& 2 v^{-2} + \log v^2 + O(v^0), ~~~~~~~ m=2/3 \\
\Gamma[-2,v^2/2] &\sim& 2 v^{-4} - 2 v^{-2} - O(\log(v)), ~~~ m=2/5
\end{eqnarray}
Thus, quadratic scaling of the structure function with velocity
is preserved even in these cases, except that for the Kolmogorov turbulence,
logarithmic correction enters the leading behaviour.

As in the case of the radial dependence (\ref{AppC:eq_dvR}), the velocity
term can be considered a special case $\gamma=0$ of the density term.
Indeed $\tilde d_\rho(0,v)$ is obtained analogous to just completed analysis
by substituting $p=1/m-\gamma/m-1/2$ and restoring $(r_0/S)^\gamma$ amplitude
factor.
\begin{eqnarray}
\tilde d_\rho(0,v) &\propto& \frac{{\bar \rho}^2 S^2}{ D_z(S)} 
\left(\frac{r_0}{S}\right)^\gamma \;
\Gamma\left(\frac{1}{2}-\frac{1-\gamma}{m}\right)
\left(\frac{v^2}{D_z(S)}\right)^
{\frac{1-\gamma}{m}-\frac{1}{2}}, 
~~~ m > 2/3 \; (1-\gamma)
\label{AppCeq:d_rhoVf1} \\
\tilde d_\rho(0,v) &\propto& \frac{{\bar \rho}^2 S^2}{ D_z(S)}
\left(\frac{r_0}{S}\right)^\gamma \;
\frac{1}{2(1-\gamma)-3m} \; \frac{v^2}{D_z(S)} ~~~~~~~~~~~,
~~~~ m < 2/3\; (1-\gamma)
\label{AppCeq:d_rhoVf2}
\end{eqnarray}
Notably, the range of $m$ where velocity scaling is sensitive to $m$
decreases for the long-range dominated density spectra, $\gamma < 0$.
For example, for Kolmogorov density spectra $\gamma=-2/3$,  we should use
the second scaling solution for all $m < 10/9$. Thus, the quadratic scaling
solution (\ref{AppCeq:d_rhoVf2}) is the one of predominant interest
in the study of such density inhomogeneities.
On the other hand, for the
short-wave dominated density spectra $\gamma > 0$ the $m$ sensitivity
is retained over much wider range and one should use expression
(\ref{AppCeq:d_rhoVf1}).
\footnote{However, for $\gamma + m/2 \ge 1$ our analysis should
be replaced by the PPV correlation function formalism.}
Exact transitional value of $m=2/3 (1-\gamma)$ points again to the presence
of the slow varying logarithmic term in the asymptotic scaling.

One may think that there is a contradiction of our present results with
those in LP00. Indeed, spectra along the v-direction obtained in LP00
were found to reflect the spectral index over the whole range of spectral
indexes studied. At the same time, structure functions of fluctuations
measured along the velocity coordinate do not have this property. 
This contradiction resolves in a very simple way. The structure functions
cannot grow faster than $v^2$. As the result they do saturate at this
value. At the same time spectra do not have such a limitations and 
properly reflect very steep spectra that arise. Therefore to identify
the power index of fluctuations along the velocity coordinate it is necessary to measure
spectra along the velocity axis. This provides a new way\footnote{Note, that
it is not necessary to have good {\it spatial} resolution to get 
turbulence if spectra along the velocity coordinate are taken.
Therefore velocity turbulence studies may become an important part of
extragalactic research. The corresponding asymptotics can be found in LP00,
Table 3.} to study turbulence that is the subject of a current
research in collaboration with Alexey Chepurnov that
we shall elaborate elsewhere. Here we shall only note that the steepness
of the spectrum prevents a useful generalization of the Spectral Correlation
Function approach (see Rosolowsky et al. 1999) to the velocity direction
(compare to a discussion in Lazarian 1999).

\section{Optically thin case: structure function approach}
In this Appendix we revisit the problem of the intergalactic turbulence
in the case of optically thin lines, which was the focus of LP00.
With the correlation function in velocity space given
by (\ref{ksigal}) we are now able to directly derive the 
asymptotic expressions 
for the correlation (or structure) functions of the intensity,
bypassing rather cumbersome calculations of the 3D power spectrum in
the velocity space.

The intensity of the emission in an optically thin line is proportional
to the density of the atoms in the velocity space
\begin{equation}
I({\bf X}_1,v_1) = \epsilon \rho_s({\bf X}_1,v_1)
\end{equation}

The mean line profile follows from eq.~(C3)
\begin{equation}
\langle I({\bf X}_1,v_1) \rangle = \frac{\epsilon \bar \rho}
{\left( 2 \pi \left[\beta + D_z(\infty)\right]\right)^{1/2}}
\int_0^S dz \; \exp\left[-\frac{(v_1-v_{gal}({\bf x}_1))^2}{2 \left[\beta+ D_z(\infty)\right]}\right]
\end{equation}

Total intensity in a velocity channel of width $\delta V$
centered at velocity $V$
and described by the shape function $W_C(v-V)$, such that
$\int dv W_C(v-V)=\delta V$ is
\begin{equation}
I_C({\bf X}_1) = \epsilon \int dv_1 \; W_C(v_1-V) \; \rho_s({\bf X}_1,v_1)
\end{equation}
The mean of the channel-integrated intensity is
\begin{equation}
\langle I_C({\bf X}_1) \rangle = \frac{\epsilon \bar \rho}
{\left( 2 \pi \left[\beta + D_z(\infty)\right]\right)^{1/2}}
\int_0^S dz \; \int dv W_C(v-V)\;
\exp\left[-\frac{(v-v_{gal}({\bf x}_1))^2}{2 \left[\beta+ D_z(\infty)\right]}\right]
\end{equation} 
For the correlation function, 
\begin{equation}
\xi_I({\bf R}) \equiv \langle I_C({\bf X}_1) I_C({\bf X}_2)\rangle =
\epsilon^2 \int\,dv_1 W_C(v_1)\int\,dv_2 W_C(v_2) 
\langle \rho_s({\bf X}_1,v_1)\rho_s({\bf X}_2,v_2) \rangle~,
\end{equation}
we shall use expression (\ref{AppB:2point_inf}) applicable to the case
of intergalactic turbulence with extended gas distribution and
coherent galactic flow with a linearized shear:
\begin{equation}
\xi_I({\bf R}) \propto \frac{\epsilon^2 {\bar\rho} ^2} {2 \pi}
\int_{-\infty}^\infty {\mathrm d}z \;
\left[1+\tilde \xi({\bf r})\right] \;
[D_z({\bf r})+2\beta]^{-1/2} \int {\mathrm d}v W(v)
\exp\left[-\frac{(v-f^{-1}z)^2}{2 (D_z({\bf r})+2\beta)}\right]
\label{AppD:xiR}
\end{equation}
Here two velocity channel windows are combined into
$W(v) = \int {\mathrm d}v_+\; W_C(v_+/2-v) W_C(v_+/2+v) $.

Equation~(\ref{AppD:xiR}) 
describes the effect of the line-of-sight projection
of 3D turbulent distributions of emitters.
Velocity kernel both modifies the projection of the density correlation function
($\xi({\bf r})$ term in the first bracket), which now exhibits different scaling
than just a column density correlation function would, and also introduces
a new phenomena - even uniform spatial distribution of emitters
following incompressible turbulent flow will be cause with fluctuating
intensity if observations are carried out in sufficiently narrow velocity 
channel. The pure velocity induced fluctuations are defined
by the first, namely, 
unity term in the first bracket. Further one, we shall discuss
two effects separately, using obvious split of the full correlation function
into two terms $\xi_I({\bf R})=\xi_v({\bf R})+\xi_\rho({\bf R})$.

As expected, the intensity integrated over the whole line contains no
information about the velocity field.
Indeed, integration with a constant window $W(v)$
over all velocity range in eq.~(\ref{AppD:xiR}) gives
\be
\xi_\rho({\bf R}) \sim \frac{1} {2 \pi}
\int_{-\infty}^\infty {\mathrm d}z \; \xi({\bf r}) \sim R^{1-\gamma}
\ee
while $\xi_v$ is reduced to a constant which describes uniform mean intensity
distribution and can be dropped.

In LP00 we presented a detailed discussion of the velocity influence on the
projected intensity statistics in optically thin lines. Here we shall
reproduce the scaling solutions at small sky separations $R$
which was shown in LP00 can be found for observations 
in {\it thin} velocity channels, using correlation function formalism
of eq.~(\ref{AppD:xiR}) as the starting point.

For the very narrow channel $W(v)=\delta(v)$ we have
\be
\xi({\bf R}) \sim \frac{1} {2 \pi}
\int_{-\infty}^\infty {\mathrm d}z \; \left[1+\tilde \xi({\bf r})\right]
[D_z({\bf r})+2\beta]^{-1/2}
\exp\left[-\frac{(f^{-1}z)^2}{2 (D_z({\bf r})+2\beta)}\right]
\ee
The asymptotic behaviour of the density term at
the small scales $R \ll (C^2 f)^{\frac{1}{2-m}}$ is obtained straightforwardly:
\begin{equation}
\xi_{\rho}({\bf R}) \sim \frac{1} {2 \pi}
\int_{-\infty}^\infty {\mathrm d}z \; \tilde \xi({\bf r})
[D_z({\bf r})+2\beta]^{-1/2}
\sim \frac{1} {2 \pi}
\int_{-\infty}^\infty {\mathrm d}z \; \tilde \xi({\bf r})/D_z({\bf r})^{-1/2}
 \sim R^{1-\gamma-m/2}~.
\end{equation}
The latter expansion, valid when $R$ is still
large enough for the turbulent dispersion to exceed the thermal one
$D({R}) > \beta $, is the one of the {\it thin} slice regime. Importantly,
the slope is modified by the term $m$, reflecting the statistical properties
of the turbulent velocity.

Asymptotic analysis of the pure velocity contribution
\begin{equation}
\xi_v({\bf R}) \sim \frac{1} {2 \pi}
\int_{-\infty}^\infty {\mathrm d}z \; 
[D_z({\bf r})+2\beta]^{-1/2}
\exp\left[-\frac{(f^{-1}z)^2}{2 (D_z({\bf r})+2\beta)}\right]
\end{equation}
is more delicate, since the integral $\int_{-\infty}^\infty {\mathrm d}z \; 
[D_z({\bf r})+2\beta]^{-1/2}$ does not converge and the exponential factor
cannot be just set to unity in the leading order. The following manipulations
lead to the correct result:

Using $D({\bf r})=C (R^2+z^2)^{m/2}$, omitting thermal term for brevity
and introducing new variable
$y=\lambda^{m/2-1} z/(R^2+z^2)^{m/4}$, $\lambda=(C^{1/2}f)^{2/2-m}$ we obtain
\begin{equation}
\xi_v({\bf R}) \sim \frac{f} {2 \pi}
\int_{-\infty}^\infty {\mathrm d}y \; 
\frac{R^2+z^2}{R^2+z^2(1-m/2)} e^{-y^2/2}
\end{equation}
where we have left $z$ as the implicit function of $y$.
Now we integration by parts
\begin{equation}
\xi_v({\bf R}) \sim \frac{f} {2 \pi}
\left\{
\left.\mathrm{erf}\left[\frac{y}{\sqrt{2}}\right]
\frac{R^2+z^2}{R^2+z^2(1-m/2)} \right|_0^\infty
-\int_{-\infty}^\infty \mathrm{d}y \;
\mathrm{erf}\left[\frac{y}{\sqrt{2}}\right]
\frac{\mathrm{d}}{\mathrm{d}y}
\frac{R^2+z^2}{R^2+z^2(1-m/2)} 
\right\}
\end{equation}
The first term is finite, while in the second one we replace the
integration variable
back to $\tilde z=z/R$ and expand the error function in power series in
$y=(R/\lambda)^{1-m/2} \tilde z/(1+\tilde z^2)^{m/4}$ keeping the
dominant first linear term.  The remaining integral is convergent 
and the scale dependent term in the correlation function
can be expressed through Gamma and Hypergeometric functions
\begin{equation}
\xi_v({\bf R}) \sim 
\frac{2}{2-m} -(R/\lambda)^{1-m/2}\; \frac{\Gamma[m/4-3/2]}{2^{3/2} \Gamma[m/4]}
{\;}_2F_1\left(\frac{3}{2},2,\frac{5}{2}-\frac{m}{4},1-\frac{m}{2}\right)~.
\end{equation}
Actually, only $m/(2-m)$ part of the second moment $\xi_v(0)=2/(2-m)=1+m/(2-m)$
describes the variance of the fluctuations of intensity, while the extra unity 
(in our normalization) comes from a nonzero mean value of $I({\bf X})$.
The structure function, insensitive to the mean value of the field, is
\begin{equation}
D_v({\bf R}) \sim 
(R/\lambda)^{1-m/2}\; \frac{\Gamma[m/4-3/2]}{2^{1/2} \Gamma[m/4]}
{\;}_2F_1\left(\frac{3}{2},2,\frac{5}{2}-\frac{m}{4},1-\frac{m}{2}\right)
\end{equation}

The Table~\ref{tab:comparison} summarizes these small $R$
asymptotics of correlation functions,
and the corresponding behaviour of the power spectra
\begin{table}[t]
\begin{displaymath}
\begin{array}{lll}
case & \xi({\bf R}) & P({\bf K}) \\
{\it thick}~slice &  R^{1-\gamma} & K^{-3+\gamma} \\   
{\it thin}~slice &  R^{1-\gamma-m/2} & K^{-3+\gamma+m/2} \\   
{\it thin}~slice, velocity & R^{1-m/2} & K^{-3+m/2}
\end{array}
\end{displaymath}  
\label{tab:comparison}
\end{table}

The correlation function formalism, developed in this Appendix, also makes
self-evident the distinction between {\it thin} and {\it thick}
channels, introduced in LP00. Direct inspection of
eq.~(\ref{AppD:xiR}) shows that one can recover {\it thin} slice behaviour
if the width of the channel window $\delta V$ combined with thermal broadening
does not exceed the turbulent velocity amplitude at the scale of consideration
$\delta V^2+2\beta\ll D(R)$ while if opposite is true the slice is {\it thick}. 
Naturally, this criterion coincides with that derived in LP00.

\section{Gaussian fluctuations: General case}

While in section~3 we attempted to be as general as possible while dealing
with correlations at small angular separations, here we deal with arbitrary
angular separations. The price that we pay for that is that we have to make
an assumption about the statistics from the very beginning. Namely,
we assume that fluctuations are Gaussian with the probability function:
\begin{eqnarray}
P=\frac{1}{2\pi C^{1/2}}
\exp\left(-\frac{1}{2C} (\delta\rho_1^2 A
-2\delta\rho_1 \delta\rho_2 \xi_s + \delta\rho^2_2 A)\right)~~~,\nonumber
\\
C\equiv A^2 - \xi_s^2,~ A\equiv \langle (\delta \rho_s)^2\rangle,
\label{probability}
\end{eqnarray}
$A$ is the dispersion  and the information about the field correlation
is contained in $\xi_s$. 

The exponent in (\ref{simplified})
 can be rewritten in terms of  fluctuations
$\exp(\alpha \rho_s )\equiv \exp(\alpha \bar \rho_s ) \exp(\alpha \delta\rho_s)$.
Then eq.~(\ref{28}) can be rewritten: 
\begin{eqnarray}
{\cal D}({\bf R})&=&\frac{\epsilon^2}{\alpha^2}
\int\, dv_1 W(v_1) \int\,dv_2 W(v_2)
\times \nonumber \\
&\times& 
e^{-\alpha (\bar \rho_{11} + \bar\rho_{12})} \left[
\langle e^{-\alpha(\delta \rho_{11} + \delta \rho_{12})} 
+ e^{-\alpha ( \delta\rho_{22} + \delta\rho_{21})}
 - e^{-\alpha ( \delta \rho_{21} + \delta \rho_{12} )}
- e^{-\alpha ( \delta \rho_{11} + \delta \rho_{22} )} \rangle\right].
\end{eqnarray}

To evaluate the averages in (\ref{dr_emiss}) with probability distribution
(\ref{probability}) we observe\footnote{The procedure of
averaging with distribution given by eq.(\ref{probability})
is straightforward (see LP00) and amounts to taking Gaussian integrals
over $y_1$ and $y_2$.} that 
\be  
\langle \exp\left[-\alpha(\delta\rho_{ii}+
\delta\rho_{jj})\right]\rangle=  
\exp\left[\frac{\alpha^2}{2}\left\langle(\delta\rho_{ii}+
\delta\rho_{jj})^2 \right\rangle \right]~~~,
\label{e3}
\ee
where quantities $\delta \rho_{ii}$ and
 $\delta \rho_{jj}$ are measured
at points ${\bf X_i}, v_i$ and ${\bf X_j}, v_j$ of the data cube.

Assuming homogeneity in the velocity distribution
we get correlations that 
depend only on the velocity difference $v$ \footnote{
A complication arises from the fact that the emission line is finite
in the velocity space and therefore the velocity distribution, in general,
not homogeneous. The finite velocity extend of the emission line
guarantees that contribution coming from $v$ much larger than the  
velocity dispersion tends to zero. An analogous effect arising
from finite boundaries of an emitting region was treated in LP00
where it was shown that for sufficiently small scales under study
the distortions introduced by boundaries are marginal.}
 between the points and
the separation between the points: 
\be
{\cal D}({\bf R})\propto \frac{\epsilon^2}{\alpha^2}
\int dv W_2(v) \left(\exp\left[\alpha^2\xi_s(0,v)\right]-\exp\left[
\alpha^2\xi_s({\bf R},v)\right]\right)~~~,
\label{largeD1}
\ee
The expansion of the exponent for small separations ${\bf R}$ results in
eq.~(\ref{dmain}) taking into account that 
$2[\xi_s(0,v)-\xi_s({\bf R},v)]=d_s({\bf R},v)-d_s(0,v) $.

\end{document}